\documentclass[10pt]{article}
\usepackage{iftex}
\usepackage{xparse}

\ifpdftex
    \usepackage[T1]{fontenc}
\else
    \usepackage{fontspec}
\fi

\usepackage{geometry}

\usepackage{amsmath}
\usepackage{amssymb}
\usepackage{amsthm}
\usepackage{xfrac}

\usepackage{xcolor}
\usepackage{graphicx}
\usepackage[colorlinks=true, citecolor=blue]{hyperref}
\usepackage{ulem}

\usepackage{wrapfig}
\usepackage{enumitem}

\newtheorem{thm}{Theorem}[section]
\newtheorem{prop}[thm]{Proposition}
\newtheorem{lemma}[thm]{Lemma}
\newtheorem{corol}[thm]{Corollary}
\newtheorem{defn}[thm]{Definition}

\newtheorem{remark}[thm]{Remark}

\newcommand{\ds}{\displaystyle}
\newcommand{\ben}{\begin{equation}}     
\newcommand{\eeqn}{\end{equation}}
\newcommand{\bey}{\begin{eqnarray}}
\newcommand{\eey}{\end{eqnarray}}

\DeclareMathOperator{\tr}{tr}
\DeclareMathOperator{\sgn}{sgn}

\begin{document}


\begin{flushleft}
{\Large
\textbf{Asymptotic dynamics in systems of two coupled quadratic maps}
}
\\
\vspace{4mm}
Anca R\u{a}dulescu$^{*,}$\footnote{Assistant Professor, Department of Mathematics, State University of New York at New Paltz; New York, USA; Phone: (845) 257-3532; Email: radulesa@newpaltz.edu}, Eva Kaslik$^{2,3}$, Alexandru Fikl$^3$\\

\indent $^1$ Department of Mathematics, SUNY New Paltz, USA

\indent $^2$ Department of Computer Science, West University of Timi\c{s}oara, Romania

\indent $^3$ Institute for Advanced Environmental Research, West University of Timisoara, Romania
\end{flushleft}


\begin{abstract}

\noindent In our previous work, we initiated the study of dynamics in networks with identical nodes that evolve under complex quadratic iterations. For these Complex Quadratic Networks, (or CQNs for short), we focused on defining the equi-M set as the natural extension of the Mandelbrot set, representing the locus of complex parameters for which the network’s critical orbit remains bounded.

This paper examines the structure and limitations of equi-M sets in two-dimensional Complex Quadratic Networks (CQNs). In particular, we aim to describe the relationship between the equi-M set and the parameter domains where the critical orbit converges to periodic attractors. The simplest case of two coupled nodes serves as a foundational testbed: its analytical tractability enables the identification of critical phenomena and their dependence on coupling, while offering insight into more general principles. The two-node case is also simple enough to allow for explicit characterization of the coupling conditions that govern transitions between synchronized and desynchronized behavior.

Using a combination of analytical and numerical methods, the study reveals that while the parameter region corresponding to an attracting fixed point closely tracks the boundary of the equi-M set near its main cusp, this correspondence breaks down for higher periods and in regions supporting coexisting attractors. These discrepancies highlight key differences between uncoupled one-dimensional dynamics and coupled systems, where equi-M sets capture only partial information about the system’s global combinatorics. Overall, the results illustrate how coupling transforms the familiar structure of the Mandelbrot set into a richer, higher-dimensional landscape, the topological and dynamical behavior of low-dimensional CQNs and point toward a sharp increase in complexity as the number of nodes grows. This work bridges classical complex dynamics with emerging questions in networked nonlinear systems and lays the groundwork for analyzing high-dimensional coupled dynamics.
\end{abstract}

\section{Introduction}

In our previous work~\cite{Ariel,Simone,gender,AC2,DTI}, we initiated the study of Complex Quadratic Networks (CQNs), systems in which each node evolves according to a complex quadratic iteration, and the nodes influence each other through a real-valued coupling matrix. CQNs extend the classical quadratic map $f_c(z)=z^2+c$ to interacting settings, allowing us to study how coupling structure alters the familiar behavior of single complex maps. As mathematical objects, CQNs exhibit a much wider range of behaviors compared to standard  one-dimensional quadratic maps. These behaviors are governed by the interplay between the properties of individual nodes, the strength of the coupling between them, and the topology of the network. For instance, studying the effect of edge placement and weight strength is of crucial interest in the context of understanding and classifying the long-term behavior of the system~\cite{iniguez2020bridging,seabrook2021evaluating,faskowitz2022edges}.Existing results for CQNs have revealed that even small adjustments in coupling can reorganize the asymptotic dynamics, revealing how structural connectivity shapes emergent behavior. This has significant implications not only for network science, but also for other fields such as physics~\cite{robinson2022physics}, biology~\cite{ma2012biological,kahn2025network}, medicine~\cite{luppi2024contributions} and engineering~\cite{whitney2004influence}, where understanding the dynamics of complex systems is a critical component.

However, many questions in network sciences are extremely difficult to address in a realistic model of a natural system, and are deemed analytically, and often also computationally, intractable. Our goal is to develop a canonical and descriptive mathematically framework that provides a systematic way to study how coupling architecture shapes the overall patterns that emerge from interaction. Rephrasing such a question around the coupled dynamics of complex iterated maps presents a mathematically tractable and canonical approach, with interesting mathematical novelty and potential for understanding natural systems.

\subsection{Combinatorics in CQNs}

More specifically, we study the properties of a general network model in which the nodes are complex quadratic maps with coupling specified by the weighted adjacency matrix $A \in \mathbb{R}^{m \times m}$~\cite{AC2,Simone,Ariel}. The overall evolution of this discrete system takes the form of an iteration in $\mathbb{C}^m$, given component-wise by
\begin{equation} \label{eq.quadratic.system}
z_k(n + 1) = \left(\sum_{j = 1}^m A_{kj} z_j(n)\right)^2 + c_k.
\end{equation}

Over the decades, the study of the dynamics of discrete iterated maps has developed into a rich research field~\cite{carleson1996complex}, with univariate quadratic polynomials taking a central position~\cite{mcmullen1994complex}. For these systems, real quadratic iterations have first been associated with cascades of period-doubling bifurcations and universality. On the other hand, complex quadratic iterations in the family $f_c(z) = z^2 + c$, with $c \in \mathbb{C}$, provide textbook recipes for creating fractal asymptotic sets. Together, they represent a simple, yet incredibly rich, family of maps that have seen many practical applications.

For univariate quadratic maps, one of the most striking results is that the orbit of the critical point captures global information on all other accessible orbits. Subsequently, one object of interest has been the set of parameters $c \in \mathbb{C}$ for which the critical orbit of $f_c$ is bounded, also known as the Mandelbrot set. It is now well-known that values of $c$ in the Mandelbrot set correspond to parameters for which the prisoner set of $f_c$ (i.e., the set of points with bounded orbits under $f_c$) forms a connected set. If $c$ is not in the Mandelbrot set, then the prisoner set is totally disconnected. This result is known as the Fatou--Julia Theorem~\cite{milnor2006dynamics}. Furthermore, the topological properties of the Mandelbrot set in the parameter plane classify all possible attracting periodic orbit combinatorics for quadratic functions in the family $f_c$, $c\in \mathbb{C}$, via a ``hyperbolic bulb'' structure. Specifically, the main cardioid of the Mandelbrot set bounds the set of parameters for which the critical  orbits converge to an attracting fixed point~\cite{brucks2004topics}. The higher-order ``hyperbolic bulbs'' correspond to the respective higher-order periodic attractor combinatorics~\cite{ivancevic2007high}.
Establishing  connections between quadratic dynamics and the topology of the Mandelbrot set has been highly nontrivial.
While many interesting properties have been established (such as connectedness~\cite{douady1984etude,douady1985dynamics} and full Hausdorff dimension~\cite{shishikura1998hausdorff}), there are open problems that persist (such as local connectedness~\cite{hubbard1992local} and density of hyperbolicity~\cite{douady1984exploring,jung2002homeomorphisms}). Even so, the Mandelbrot set remains a staple in discrete dynamical systems and a canonical representation for more general phenomena.

In our work on CQNs, we track asymptotic behavior of multi-dimensional orbits, and use it to quantify how systemic dynamics emerge under different specifications for network architecture. For any given network with identical nodes $c$, we defined the extension of the Mandelbrot set as the postcritically bounded locus for $c$ (and called it the equi-M set of the network). We noted that equi-M sets are combinatorially more complex than the traditional Mandelbrot set, due to the interactions between the multiple nodes of the network. In particular, since the system is no longer generated by a single unimodal map, we can no longer expect that the critical orbit encapsulates the behavior of all orbits of the system. In our previous work~\cite{Ariel,Simone,AC2,DTI}, we explored post-critical dynamics for CQNs
using the equi-M, as well as node-wise versions of it.

Our initial observations suggested that the equivalence of the post-critically bounded and connectedness loci does not hold in the same form for CQNs as it did for univariate iterated maps, where the Mandelbrot set provides an atlas of all periodic attractor regions, and implicitly for the behavior of prisoner sets across the quadratic family. In our preliminary work~\cite{Simone}, we suggest potential extensions to capture the fact that the critically bounded locus for CQNs is only related in a weaker sense to the connectedness locus in $\mathbb{C}^m$. Here, we revisit the question and show the mechanisms that shape this weaker version of the Fatou--Julia relationship through the underlying combinatorics of CQNs. For example, we expect that there may be CQNs which have periodic attractors, but for which the critical orbit escapes, breaking one of the most basic theorems for the dynamic family defined by univariate quadratic iterated maps. 

\subsection{Dynamics of two coupled maps}

In this paper, we focus on setting the groundwork for a few questions that go along these lines. One consists of finding an effective atlas that would extend the idea of hyperbolic components as a combinatorial bookkeeping method. This organization would have to go beyond the traditional partition in hyperbolic bulbs, and would have to allow for more complex possibilities, such as parametric regions in which multiple period attractors coexist. We study to what extent the equi-M set offers a roughly approximate (but computationally more inexpensive) representation of this atlas, and which features of this set remain useful for characterizing and classifying the asymptotic behavior of the system. Finally, we want to understand how behavior differs between individual nodes, and how boundedness of the different node-wise components of the critical orbit contribute to shaping the intersection (which is the equi-M set).

We focus our study on systems consisting of two coupled nodes, as a first step to investigate the complexity of network combinatorics and the relationship between the combinatoric atlas and the shape of the equi-M set. We focus on two aspects of the equi-M set: topology and synchronization. First, we use analytical and computational tools to identify low-period attractors, and sketch the skeleton of the combinatoric atlas for different types of two-dimensional systems. Then, we compare the geometry of these structures with that of the corresponding equi-M sets, generated numerically. Second, we examine equi-M sets from the perspective of node-wise contributions, identifying the factors that make nodes ``synchronize'', i.e. exhibit the same parameter locus for which their individual components of the critical orbit are bounded. In our previous work~\cite{Simone,AC2}, we determined that network nodes may cluster into groups that have the same node-wise equi-M set, based on the architectural profile of the network. However, the contribution of specific hard-wired factors to this synchronization profile remained unclear in large oriented networks. The study of two-dimensional coupled systems allows us to identify possible basic mechanisms for synchronization in this simple case, that can be further extended to more general systems.

More generally: in our previous work, we have approached CQNs from a network science standpoint~\cite{AC2,gender}, and have focused primarily on the contribution of the network architectural aspect to the emerging dynamics. In particular, we worked to identify the contribution of the network coupling to the topological properties of the equi-M sets, and to the grouping and shapes of the synchronization clusters. However, in order to have nontrivial architecture, the network has to be reasonably large. This leap to access higher-dimensional, interesting ``connectivity patterns'' omitted a specific analysis of two-dimensional networks, which can only have virtually trivial architecture schemes between two nodes. This case is nonetheless extremely important to our analytical understanding of how coupling affects the equi-M sets and their synchronization. In this paper, we focus precisely on this aspect and embrace a bottom-up, constructive approach. In future work we will build upon results on interaction between two nodes taken in isolation, to determine how mutual coupling acts when it is more generally embedded into higher-dimensional networks.

The paper is organized as follows: In Section~\ref{combinatorics}, we introduce the general concepts and notation for systems of two coupled maps, including the definition of equi-M set in this case. In Subsection~\ref{escape_rad} we establish the existence of escape radius for two coupled maps (a crucial piece supporting reliability of numerical simulations based on direct iterations of the system). In Subsection~\ref{cardioid} we define and describe the ``main equi-cardioid'' of the equi-M set in this context, in analytic and numerical terms, for two-dimensional systems in general and for a few example families. Regions with higher-order period attractors are computed numerically and illustrated in Subsection~\ref{higher_order}. Section~\ref{synchronization} describes synchronization in coupled systems and analyses phase transitions between clustering and de-clustering of the two nodes. The results are then reviewed and contextualized in the Discussion section.

\section{Combinatorics of coupled maps}
\label{combinatorics}

We consider a system of two coupled complex variables $(z_1, z_2)$ that evolve in discrete time, according to quadratic complex dynamics~\eqref{eq.quadratic.system} for identical complex parameters $c_1=c_2=c$, with linear coupling specified by a matrix $A \in \mathbb{R}^{2 \times 2}$. The two-dimensional map describing the coupled dynamics is given by:
\begin{equation} \label{mother_sys}
\begin{cases}
z_1(n+1) = [a z_1(n) + b z_2(n)]^2 + c, \\
z_2(n+1) = [d z_1(n) + f z_2(n)]^2 + c,
\end{cases}
\end{equation}
and $c \in \mathbb{C}$. In this special case, for the sake of simplicity, we denote the elements of the matrix $A$ by $a_{11} \equiv a, a_{12} \equiv b, a_{21} \equiv d$ and $a_{22} \equiv f$. We refer to this system as a 2D-CQN to differentiate it from the general $m$-dimensional case considered in~\cite{AC2,Simone}. For this family of maps, the equi-M set is defined as follows:

\begin{defn} \label{def:equi_m_set}
The equi-M set of 2D-CQN~\eqref{mother_sys} is defined as
\[
\mathcal{M}_{2} = \{
    c \in \mathbb{C} \mid
    |z_1(n)|^2 + |z_2(n)|^2 < \infty, \forall n \in \mathbb{N},
    \text{where } z_1(0) = z_2(0) = 0
\}.
\]
\end{defn}

As in the case of univariate quadratic maps, the equi-M set is defined as the subset of the parameter space where the critical orbit remains bounded in the $\ell_2$ norm. This represents only a slice of a higher-dimensional parameter object $(c_1,c_2)\in \mathbb{C}^2$, and has a structure that depends on the coupling between nodes. When discussing the geometry of the equi-M set and its relationship to the combinatorics of~\eqref{mother_sys}, we also define the following two extensions of traditional objects from the study of univariate iterated maps.

\begin{defn} \label{eq:pseudo_bulb}
The combinatorial region $\mathcal{C}_2^k$, for $k \geq 1$, is the region of $c\in \mathbb{C}$ for which the system has an attractor of period $k$. In particular, we refer to $\mathcal{C}_2^1$ as the main equi-cardioid of the system. The pseudo-bulb $\mathcal{B}_2^k$ of period $k \geq 1$ is the region of $c \in \mathbb{C}$ where the critical point converges to an attractor of period $k$.
\end{defn}

While $\mathcal{B}_2^k \subseteq \mathcal{C}_2^k$, we note that this inclusion may be strict or become equality, depending on the system. The critical orbit no longer encompasses the global fate of the system, as it did in single iterated maps, hence $\mathcal{B}_2^k$ and $\mathcal{C}_2^k$ are distinct objects. In fact, we will show that the combinatorial regions $\mathcal{C}_2^k$ are not necessarily disjoint. This is due to the fact that periodic attractors may coexist in coupled systems, leading to a more complex combinatorial partition in the parameter plane. These properties and relationships are further examined in the following sections.

\subsection{Escape radius}
\label{escape_rad}

In this section, we show that any 2D-CQN~\eqref{mother_sys} has an escape radius under minimal assumptions on the entries of the connectivity matrix $A$. First, we prove the following lemma:


\begin{lemma} \label{case1}
Let the connectivity matrix $A$ be such that $\Delta = |a f| - |bd| \neq 0$ and $(z_1(0), z_2(0))$ arbitrary. Then, there exist a large enough $M > 0$ and an $n \ge 0$, such that
\[
\max\{|z_1(n)|, |z_2(n)|\} > M \implies \max\{|z_1(n + 1)|, |z_2(n + 1)|\} > 2 M.
\]

In particular, this is true for the critical orbit $z_1(0) = z_2(0) = 0$ that describes the
equi-M set.
\end{lemma}

\begin{proof}
Assuming that $\Delta = |af | - |bd | \neq 0$, we define
\begin{equation} \label{case1.m1m2}
M_1 = K_1 + \sqrt{K_1 (K_1 + |c|)}
\text{ and }
M_2 = K_2 + \sqrt{K_2(K_2 + |c|)},
\end{equation}
where
\[
K_1 = \left(\frac{|b| + |f|}{|\Delta|}\right)^2
\text{ and }
K_2 = \left(\frac{|a| + |d|}{|\Delta|}\right)^2.
\]

Let $M > \max\{M_1, M_2\}$. To simplify the notation, we abbreviate $z_1(n)$ and $z_2(n)$ to $z_1$ and $z_2$ for the rest of the proof, when there is no danger of confusion. Suppose that, for some $n \geq 0$, we have $|z_1| > M$ or $|z_2| > M$. We will show that $|z_1(n + 1)| > 2 M$ or $|z_2(n + 1)| > 2 M$ using a standard proof by contradiction. Suppose that both $z_1(n + 1) \leq 2M$ and $z_2(n + 1) \leq 2M$. Then, starting from $|z_1(n + 1)| \le 2 M$, we have that
\[
2 M
\geq |z_1(n + 1)| = |(a z_1 + b z_2)^2 + c|
\geq |a z_1 + b z_2|^2 - |c|
\geq (|a z_1| - |b z_2|)^2 - |c|,
\]
by repeated use of the reverse triangle inequality. Applying the analogous steps to $|z_2(n + 1)| \le 2 M$, we obtain
\[
2 M
\geq |z_2(n + 1)| = |(d z_1 + f z_2)^2 + c|
\geq |d z_1 + f z_2|^2 - |c|
\geq (|d z_1| - |f z_2|)^2 - |c|,
\]
Calling $\xi_1=|a z_1| - |b z_2|$ and $\xi_2 = |d z_1| - |f z_2|$, for convenience, we obtain that
\begin{equation} \label{case1.proof1}
    |\xi_1| \leq \sqrt{2M + |c|}
    \quad \text{and} \quad
    |\xi_2| \leq \sqrt{2M + |c|}.
\end{equation}

Using the inequalities from \eqref{case1.proof1}, we have that
\begin{equation*}
\big||f| \xi_1 - |b| \xi_2\big|
    = \big|(|a f| - |b d|) z_1\big|
    \leq |f| |\xi_1| + |b| |\xi_2|
    \leq (|b| + |f|) \sqrt{2M + |c|}.
\end{equation*}

This result provides a bound on $z_1$, under the assumption that $\Delta \ne 0$, i.e.
\begin{equation} \label{ineq_z1}
|z_1| \leq \frac{|b| + |f|}{|\Delta|} \sqrt{2 M + |c|} = \sqrt{K_1 (2 M + |c|)}.
\end{equation}

We are looking for a condition on $M$ that implies
$\sqrt{K_1 (2 M + |c|)} < M$. Squaring both sides, this is equivalent to requiring that $h_1(M) = M^2 - K_1 (2 M +|c|) > 0$. We can guarantee $h_1$ remains positive by taking $M > M_1$, the larger of its two roots. Hence, if $M>M_1$, we have that $|z_1| \leq M$. Analogously, we obtain
\begin{equation*}
   |z_2| \leq \sqrt{K_2 (2 M + |c|)},
\end{equation*}

which in turn is smaller than $M$, $M > M_2$, where $M_2$ is the larger of the roots of $h_2(M) = M^2 - K_2 (2 M + |c|)$ (as defined in~\eqref{case1.m1m2}). Since $z_1$ and $z_2$ cannot be simultaneously smaller than $M$, the contradiction follows.
\end{proof}

\begin{thm} \label{thm.escape.radius}
Let the connectivity matrix $A$ be such that $\Delta = |a f| - |bd| \neq 0$. Then $M = \max\{M_1, M_2\}$ is an escape radius for~\eqref{mother_sys}, where $M_1$ and $M_2$ are given by~\eqref{case1.m1m2}.
\end{thm}

Theorem \ref{thm.escape.radius} provides an escape radius for \eqref{mother_sys}
for the case when the matrix $A$ is not singular. However, we can extend this to
the case when $a f = b d$ (and analogously for $a f = - b d$). We analyze all the possible cases below.

\begin{description}
\item[Case 1 $(a, b \neq 0)$.]
Since $a f = b d$, we can define $k = \frac{d}{a} = \frac{f}{b}$ and consider a fixed iteration $n$, for which $|z_1(n)| > M$ or $|z_2(n)| > M$. It follows that $|a z_1 + b z_2| > g M$ or $|a z_1 - b z_2| > g M$, for $g = \min\{|a|, |b|\}$. Indeed, assume by contradiction that both $|a z_1 + bz_2| \leq g M$ and $|a z_1 - b z_2| \leq g M$. Then
\begin{equation*}
2|az_1 | \leq |az_1 + bz_2 | + |az_1-bz_2 | \leq 2gM < 2 |a | M
\end{equation*}
implying that $|z_1| \leq M$. Analogously,
\begin{equation*}
2|bz_1 | \leq |az_1 + bz_2 | + |bz_2- az_1 | \leq 2gM < 2 |b | M
\end{equation*}
implies that $|z_2| \leq M$. Since we cannot simultaneously have $|z_1| \leq M$ and $|z_2| \leq M$, the initial assumption must be false. Then, if we assume that $|a z_1 + b z_2| > g M$, we have that
\[
\begin{aligned}
& a z_1(n + 1) + b z_2(n + 1) = (a + b k^2) (a z_1 + b z_2)^2 + c (a + b) \\
\implies & |a z_1(n + 1) + b z_2(n + 1)| \geq g^2 M^2 |a + b k^2| - |c (a + b)|.
\end{aligned}
\]

\begin{description}
    \item[Case (i).] If $a + b k^2 \neq 0$, we can consider the quadratic function $f(M) = g^2M^2 |a+bk^2| -2(|a | + |b |)M - |c(a+b) |$. The larger root $M_3 > 0$ of this function is defined explicitly as:
    \begin{equation*}
    M_3 = \frac{|a| + |b| + \sqrt{(|a| + |b|)^2 + g^2 |c| (|a| + |b|) |a + b k^2|}}{g^2 |a + b k^2|}.
    \end{equation*}

    If $M > M_3$, then $f(M) > 0$, and $|a z_1(n+1) + b z_2(n+1)| > 2(|a| + |b|)M$. It follows that either $|z_1(n + 1)| > 2M$ or $|z_2(n + 1)| > 2M$. Therefore,
    $M = M_3$ is an escape radius for \eqref{mother_sys}.

    \item[Case (ii).] If $a + b k^2 = 0$, then $a = -b k^2$, $d = -b k^3$ and $f = b k$. Therefore, the system becomes
    \[
    \begin{cases}
    z_1(n+1) = b^2 (-k^2 z_1 + z_2)^2 + c, \\
    z_2(n+1) = b^2 k^2 (-k^2 z_1 + z_2)^2 + c.
    \end{cases}
    \]

    Multiplying the first equation by $-k^2$ add adding it to the second equation,
    we obtain that the quantity $-k^2 z_1 + z_2 = c (1 - k^2)$ remains constant
    throughout the iterations. If $k\neq 0$, this implies that $(z_1, z_2)$ will never escape. If $k=0$, then $a=d=f=0$, and the system is trivial.
\end{description}

\item[Case 2 $(a = 0, b \ne 0)$.]
Then, necessarily, $d = 0$ and the system becomes
\[
\begin{cases}
z_1(n+1) = (bz_2)^2 + c, \\
z_2(n+1) = (fz_2)^2 + c.
\end{cases}
\]

In this case, we can see that the second equation is decoupled. If $f \neq 0$, one can make the change of variables $\xi = f^2 z_2$, and rewrite the equation as $\xi(n+1) = \xi^2 + c_f$, where $c_f = f^2 c$. It follows from the traditional theory on escape radius for single quadratic maps that $M = \max\{2 / f^2, |c|\}$ is an escape radius for $z_2$. Hence the whole system has an escape radius, since $z_1$ only depends on $z_2$. If $f = 0$, the system is trivial as both $z_2$ and $z_1$ are constant.

\item[Case 3 $(a \ne 0, b = 0)$.] Then necessarily $f = 0$ and the system becomes
\[
\begin{cases}
z_1(n + 1) = (a z_1)^2 + c, \\
z_2(n + 1)=(d z_1)^2 + c.
\end{cases}
\]

The proof is analogous to Case 2.

\item[Case 4 $(a = b = 0)$.]
Then $z_1 = c$, for all $n \geq 0$ (constant sequence) and $z_2(n + 1)= (d c + f z_2)^2 + c$. With the change of variable $\xi = f^2 z_2 + d c$, the second equation again reduces to a standard quadratic map $\xi(n + 1) = \xi^2 + c (f^2 + d)$. The
corresponding escape radius is $M = \max\{2 / (f^2 + d), |c|\}$.
\end{description}

From the above analysis, we have that in Cases 2, 3 and 4, the 2D-CQN~\eqref{mother_sys} becomes degenerate by reducing to the dynamics of a single quadratic map. However, Case 1 implies that the escape radius property remains valid when $|a f| = |b d|$, as long as $a + b k^2 \neq 0$. As a counter-example, we can take the matrix $A$ satisfying $a = 1$, $b = -1$, and $k = 1$. We can see that initial conditions with arbitrarily large $z_1(0) = z_2(0)$ immediately collapse to $z_1(n) = z_2(n) = c$, for all $n \geq 1$. Therefore, there is no radius $M > 0$ past which orbits automatically escape to infinity.

\vspace{3mm}
\noindent {\bf Note:} Computational approaches to identifying whether orbits are bounded lead to a somewhat more general and difficult question: whether (or for which networks) it is possible that, given any large $M$, there exist $c$ values for which the critical orbit can grow larger than $M$, but remain asymptotically bounded. This can inform us for what types of networks one can test computationally whether certain values of $c$ can be unquestionably excluded from the equi-M set, based on a finite number of iterations (in the same manner as in the single map case). Investigating this statement is not within the scope of this paper.

\subsection{Stable fixed points and the main equi-cardioid}
\label{cardioid}

In this section, we examine the main equi-cardioid $\mathcal{C}^1_2$ for 2D-CQNs, i.e. the region in the complex plane where the system has at least one stable fixed point.
A fixed points $(z_1^*, z_2^*)$ of \eqref{mother_sys} must satisfy
\[
\begin{cases}
z_1^* = (a z_1^* + b z_2^*)^2 + c, \\
z_2^* = (d z_1^* + f z_2^*)^2 + c.
\end{cases}
\]

In the following analysis, we first make the change of variables $u_1^* = a z_1^* + b z_2^*$ and $u_2^* = d z_1^* + f z_2^*$. In these coordinates, the fixed points $(u_1^*, u_2^*)$ must satisfy
\begin{equation} \label{eq.fixed.point}
\begin{cases}
u_1^* = a (u_1^*)^2 + b (u_2^*)^2 + c (a + b), \\
u_2^* = d (u_1^*)^2 + f (u_2^*)^2 + c (d + f).
\end{cases}
\end{equation}

To construct the main equi-cardioid, we additionally require that, for at least one of the fixed points, the Jacobian matrix
\[
J =
\begin{bmatrix}
2 a u_1^* & 2 b u_2^* \\
2 d u_1^* & 2 f u_2^*
\end{bmatrix}
\]
has both eigenvalues within the unit disk. Equivalently, we require that both roots of the characteristic equation
\begin{equation} \label{eq.char}
\lambda^2 - 2 (a u_1^* + fu_2^*) \lambda + 4 \delta u_1^* u_2^*=0,
\end{equation}
where $\delta = \det(A) = af - bd \ne 0$, are within the unit disk. The following lemma is a consequence of the classical Schur-Cohn stability test \cite{henrici1974computational} and provides the necessary and sufficient conditions for roots of the polynomial to be in the unit disk based on its coefficients.

\begin{lemma}\label{lem.quadratic.poly}
Let $\alpha, \beta\in\mathbb{C}$. Both roots of the quadratic polynomial $x^2-2\alpha x+\beta$ belong to the open unit disk if and only if
\begin{equation} \label{ineg.lemma}
    |\beta| < 1
    \quad \text{and} \quad
    2 |\alpha-\overline{\alpha}\beta| < 1-|\beta|^2.
\end{equation}
\end{lemma}

We assume that the matrix $A$ is non-singular and, without loss of generality, that $a+b\neq 0$. We remark that eliminating $c$ from system \eqref{eq.fixed.point} gives
\begin{equation} \label{eq.u1.u2}
    (d + f) u_1^* - (a + b) u_2^* = \delta [(u_1^*)^2 - (u_2^*)^2].
\end{equation}

Based on \eqref{eq.u1.u2} and applying Lemma \ref{lem.quadratic.poly} to \eqref{eq.char} with $\alpha = a u_1^* + f u_2^*$ and $\beta = 4 \delta u_1^* u_2^*$, we obtain the following description of the main equi-cardioid.

\begin{prop}\label{prop.cardioid}
Let $a, b, d, f \in \mathbb{R}$ such that $\delta = a f - b d \ne 0$. Consider the region $\mathcal{R}$ of points $(u_1, u_2)\in\mathbb{C}^2$, which simultaneously verify
\[
\begin{cases}
    (d + f) u_1 - (a + b) u_2 = \delta (u_1^2 - u_2^2), \\
    4|\delta u_1u_2| < 1, \\
    2|a u_1 + f u_2-4\delta(a|u_1|^2u_2+fu_1|u_2|^2)| <1 - 4|\delta u_1 u_2|^2.
\end{cases}
\]
The set of complex values $c$ for which the coupled system of quadratic maps~\eqref{mother_sys} has at least one stable fixed point (i.e. the main equi-cardioid) is given by $g(\mathcal{R})$, where
\[
g(u_1,u_2)=\frac{u_1-au_1^2-bu_2^2}{a+b}.
\]
\end{prop}

Below, we provide specific examples for the main equi-cardioid of~\eqref{mother_sys} in particular cases of interest. We show a standard
feed-forward network in Section \ref{sc.cardioid.feedforward} and a coupled
network with equal row sum in Section \ref{sc.cardioid.rowsum}. Under some
assumptions, these families of 2D-CQNs allow direct computation of the main
equi-cardioid. The general case can be determined numerically
from the conditions provided in Proposition \ref{prop.cardioid}.

\subsubsection{Example 1: Feed-forward networks}
\label{sc.cardioid.feedforward}

Consider the family of feed-forward two-dimensional coupled maps given by $b = 0$ and $f > 0$, additionally requiring $a = d = 1$ (which allows computations to remain tractable). In other words:
\begin{equation}  \label{family1}
\begin{aligned}
z_1(n+1) & = z_1^2 + c \\
z_2(n+1) & = (z_1 + f z_2)^2 + c
\end{aligned}
\end{equation}

In Appendix A, we show that the boundary of the region in the $c$-plane where this two-dimensional system has at least one stable fixed point is given by the following polar curve, defined parametrically as:
\begin{equation}
\cos \theta = \frac{1-12\rho^2-4\rho^4}{4\rho(4\rho^2-1)},\quad \text{ for } \frac{3-\sqrt{7}}{2} \leq \rho \leq \frac{-1+\sqrt{3}}{2}
\tag{S1$'$} \label{S1'}
\end{equation}

\begin{figure}[htbp]
    \centering
    \includegraphics[width=\linewidth]{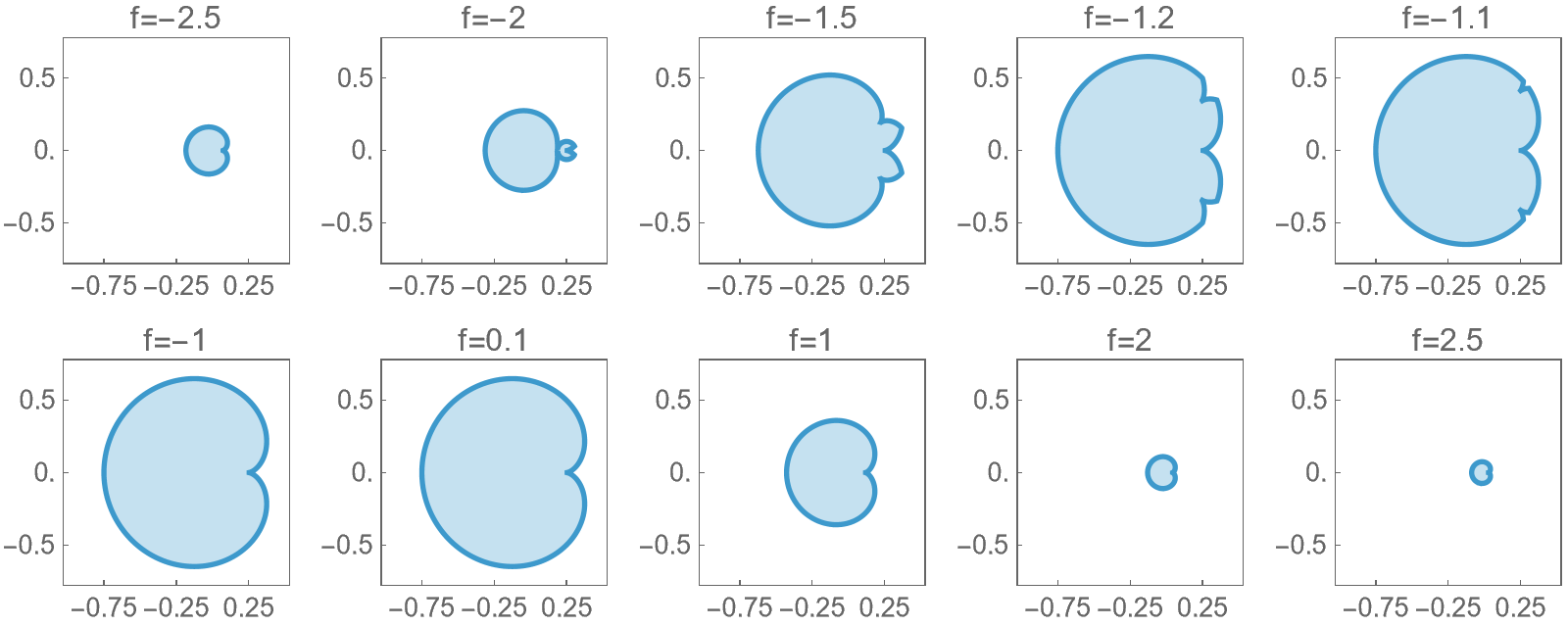}
    \caption{{\bf Main equi-cardioids} for the feed-forward case  with the coupling matrix $A=\begin{bmatrix}
        1 & 0 \\ 1 & f
    \end{bmatrix}$ for increasing values of $f$, as labeled.}
\label{fig:cardio.feedforward.1}
\end{figure}

\subsubsection{Example 2: Equal row sum networks}
\label{sc.cardioid.rowsum}

Consider the family of coupled maps subject to the condition $a + b = d + f = \omega$. Even with this simplifying assumption, the analysis becomes intractable quickly, as shown in Appendix B. The analysis for special subfamilies is more approachable. We present two interesting special cases below.

\noindent \textbf{{\it Example 2(a).}} We first consider the feed-forward case: $b = 0$. Hence, $a = \omega$, $d = 2 \omega - \tau$ and $f = \tau - \omega$. Let us denote $\rho = \tau / \omega - 1$. The characteristic equation \eqref{eq.char} simplifies to:
\[
\lambda^2 - 2(a u_1 + f u_2) \lambda + 4 a f u_1 u_2 = 0,
\]
and hence, the roots are $\lambda_1 = 2 a u_1 = 2 \omega u_1$ and $\lambda_2 = 2 f u_2 = 2(1 - (\tau - \omega) u_1) = 2 (1 - \rho \omega u_1)$, where $u_1$ is a root of
\[
\omega u_1^2 - u_1 + c \omega = 0.
\]

The boundary of the cardioid is given by either $|\lambda_1| = 1$ and $|\lambda_2| \leq  1$, or $|\lambda_2| = 1$ and $|\lambda_1| \leq 1$.

On the one hand, if $\lambda_1 = e^{i\theta}$, then $u_1 = e^{i\theta} / (2 \omega)$ and we must require that
\[
\left|2 - \rho e^{i\theta}\right| \leq 1
\iff
4 \rho \cos\theta \geq 3 + |\rho|^2
\implies 1 \leq |\rho| \leq 3.
\]

In this case, the parametric equation of a part of the boundary of the cardioid is
\begin{equation}\label{eq.c.1}
c = \frac{u_1 (1 - \omega u_1)}{\omega}
  = \frac{e^{i \theta} (2 - e^{i \theta})}{4 \omega^2},
\quad \text{where }
\sgn(\rho) \cos \theta \geq \frac{3 + |\rho|^2}{4 |\rho|}.
\end{equation}

On the other hand, if $\lambda_2 = e^{i \theta}$, then $u_2 = \frac{e^{i \theta}}{2(\tau - \omega)} = \frac{e^{i\theta}}{2 \rho \omega}$ and hence $u_1 = \frac{2 - e^{i \theta}}{2 \rho \omega}$. We must ensure that
\[
\left|2 - e^{i \theta}\right| \leq |\rho|
\iff \cos \theta \geq \frac{5 - |\rho|^2}{4}
\implies |\rho| \geq 1.
\]

In this case, we have the parametric equation
\begin{equation}\label{eq.c.2}
c = \frac{u_1 (1 - \omega u_1)}{\omega}
  = \frac{2 - e^{i \theta}}{2 \omega (\tau - \omega)}
    \left(1 - \frac{\omega(2 - e^{i \theta})}{2 (\tau - \omega)}\right)
  = \frac{(2 - e^{i\theta})(2 \rho - 2 + e^{i \theta})}{4\rho^2\omega^2},
\quad \text{where } \cos\theta \geq \frac{5 - |\rho|^2}{4}.
\end{equation}

\begin{figure}[ht!]
    \centering
    \includegraphics[width=\linewidth]{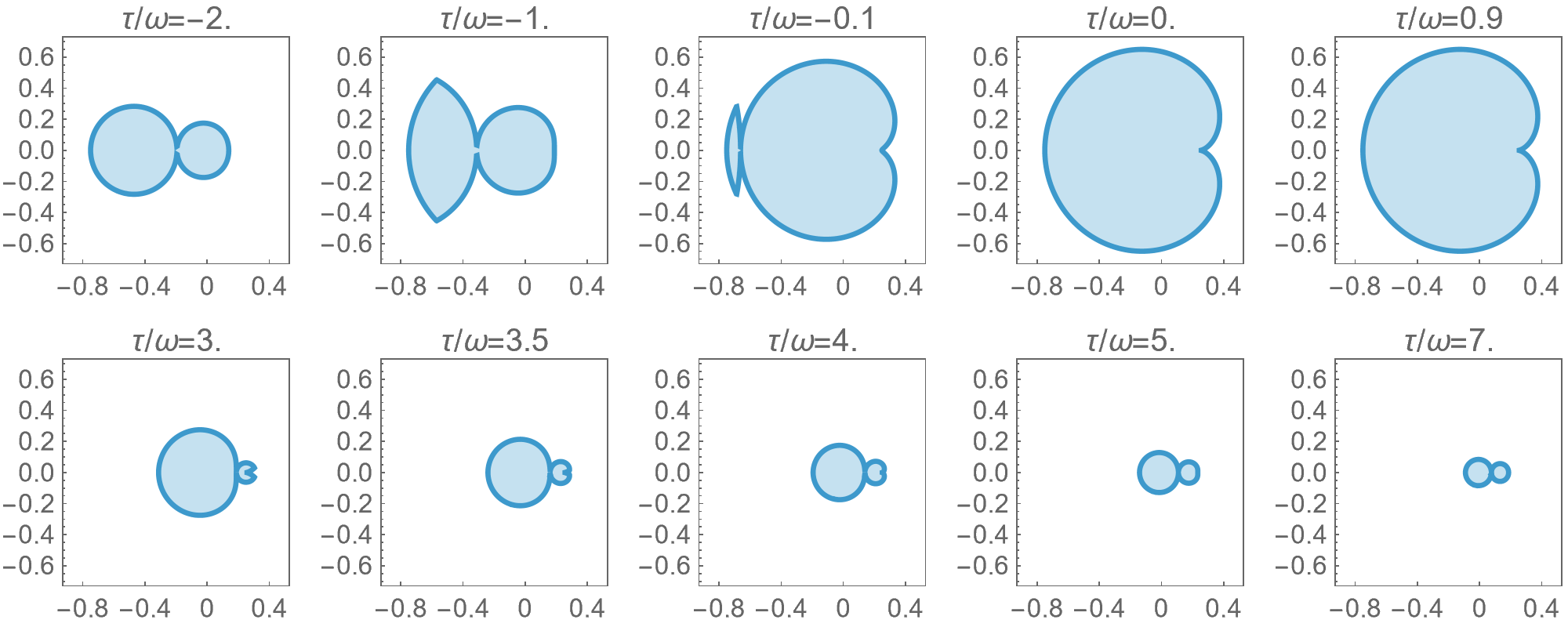}
    \caption{{\bf Main equi-cardioids} for the feed-forward case  with the coupling matrix $A=\begin{bmatrix}
        \omega & 0 \\
        2\omega-\tau & \tau-\omega
    \end{bmatrix}$ (with equal row sum $\omega$) for increasing values of $\omega$, as labeled. }
\label{fig:cardio.feedforward}
\end{figure}

In conclusion, if $|\rho| \geq 1$, the cardioid is bounded by the union of the parametric curves given by \eqref{eq.c.1} and \eqref{eq.c.2}. On the other hand, if $|\rho| < 1$, the cardioid boundary is given by the conditions derived in Appendix B, Case I, and hence, the cardioid is defined  by \eqref{ineq.cardio.1}. These two cases can be visualized in Figure \ref{fig:cardio.feedforward}.

\begin{figure}[ht!]
    \centering
    \includegraphics[width=\linewidth]{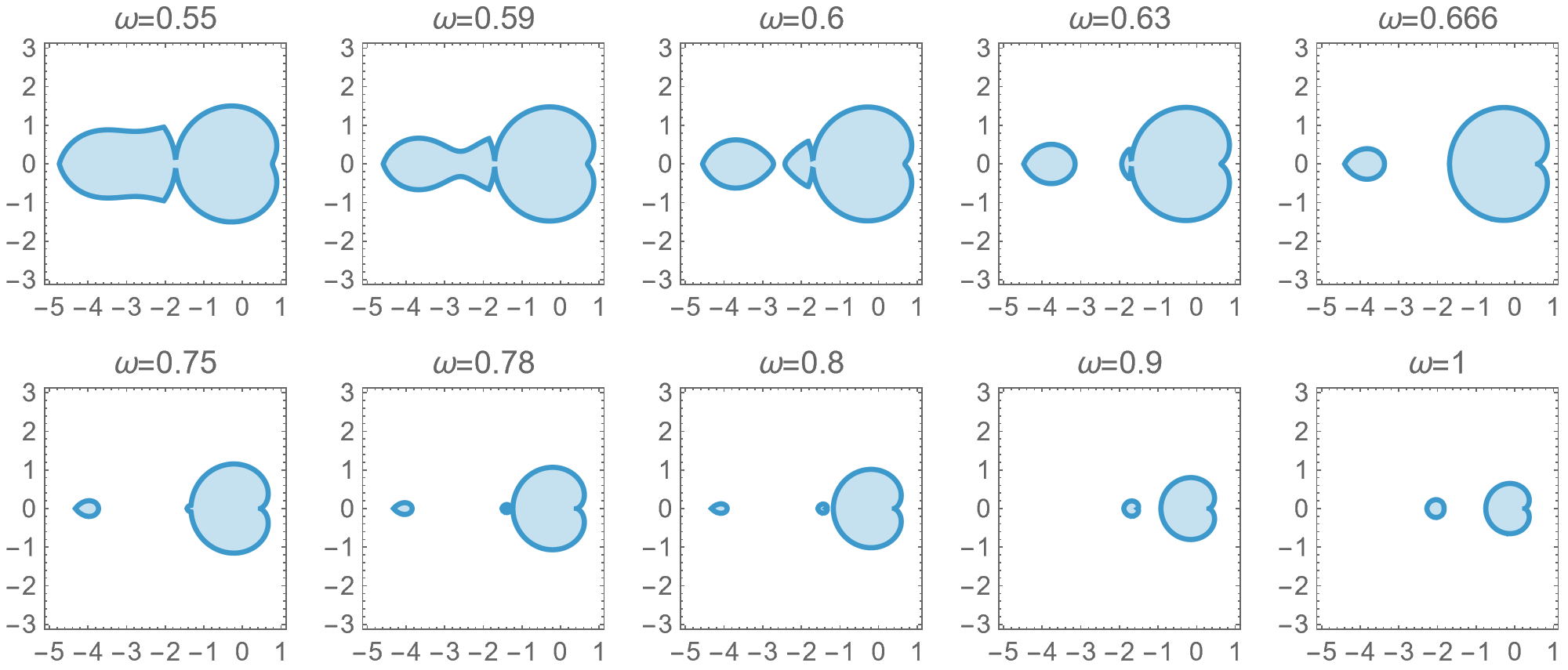}
    \caption{{\bf Main equi-cardioids} for coupled quadratic maps with the coupling matrix $A=\begin{bmatrix}
        \omega/2 & \omega/2 \\ 1 & \omega-1
    \end{bmatrix}$ for increasing positive values of $\omega$, as labeled.}
    \label{fig:cardio.w}
\end{figure}

\vspace{5mm}
\noindent
\textbf{{\it Example 2(b).}} Considering $a = b = \omega / 2$, $d = 1$ and $f = \omega - 1$, we obtain more complicated cardioid structures shown in Figure \ref{fig:cardio.w}. It is worth noting that in some cases (e.g. $\omega=0.8$), the cardioid is the union of three disjoint connected components. We also note an interesting property of the maps in this family. Due to the fact that the coupling matrix $A$ has equal row sum $\omega$, the symmetric subspace $z_1 = z_2$ is an invariant set, and the dynamics on this subspace reduces to the one-dimensional quadratic map $z \mapsto (\omega z)^2+c$. For $c = -2 / \omega^2$, this quadratic map is topologically conjugate, through the linear homeomorphism $\phi(z) = -\omega^2 z/4+1/2$, to the logistic map $z \mapsto 4 z (1 - z)$, which is known to be chaotic. Therefore, the trajectory of the 2D quadratic system starting from $(0, 0)$ is also chaotic. However, the 2D system has non-symmetric asymptotically stable fixed points for certain ranges of $\omega$ (the intervals $0.674942 < \omega < 0.737465$ and $0.967341 < \omega < 1.0212$ were found by our numerical computations to be such examples).

\subsection{Higher order periods and pseudo-bulb structure}
\label{higher_order}

We aim to classify higher order periodic combinatorics for 2D-CQNs. We do this with the understanding that coupled nodes are expected to have significantly richer behavior in $\mathbb{C}^2$ than single map iterations have in the complex plane, thus generating a much more complex atlas for potential combinatorics.

\begin{figure}[h!]
\centering
\fbox{\includegraphics[width=0.4\textwidth]{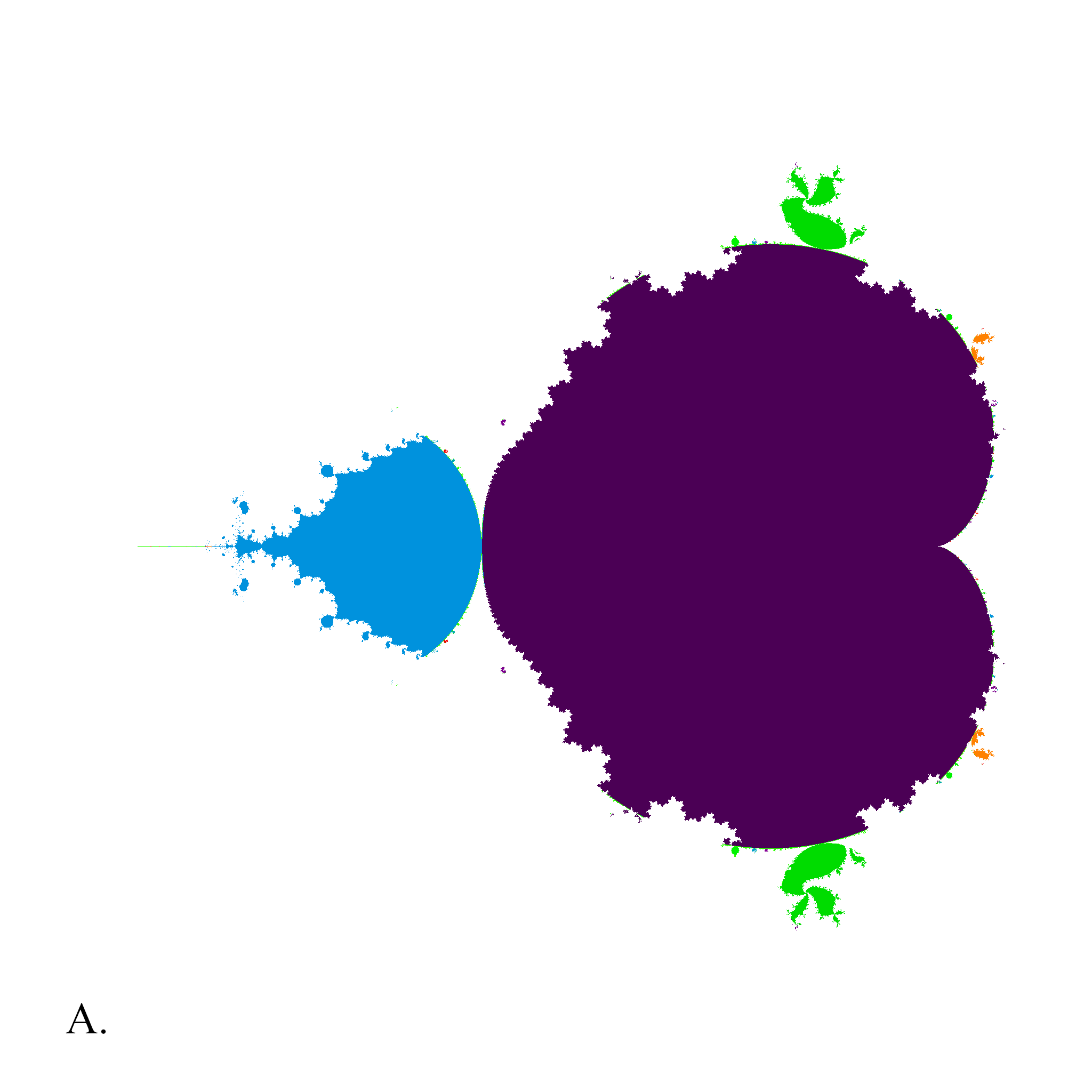}}
\quad \quad
\fbox{\includegraphics[width=0.4\textwidth]{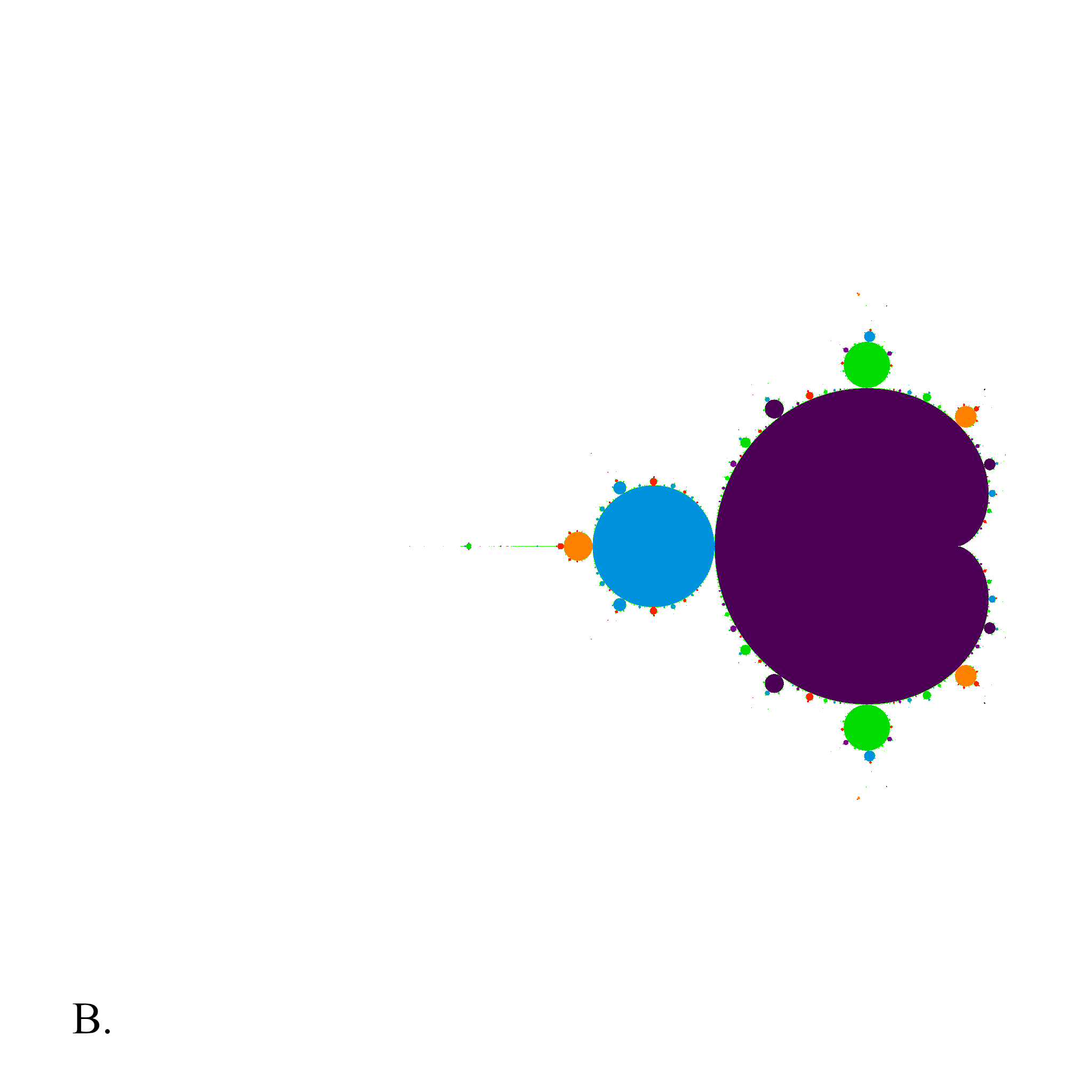}}
\caption{{\bf Post-critically periodic regions ${\cal C}_2^k$, for periods up to k=20,} are shown as subsets of the equi-M set for two example networks. \textbf{A.} $A=\begin{bmatrix}
        1 & 0 \\
        1 & 1
    \end{bmatrix}$ (Example 1 for $a=d=f=1$,$b=0$); \textbf{B.} $A=\begin{bmatrix}
        0.4 & 0.4 \\
        1 & -0.2
    \end{bmatrix}$ (Example 2 for $\omega=0.8$). The colors correspond to the critical orbit being attracted respectively to a fixed point (purple region), period two orbit (cyan region), period three orbit (green regions), period four orbit (orange regions), and so on, up to period $k=20$. }
\label{higher_periods1}

\end{figure}

To start with, we used numerical simulations to understand to what extent the equi-M set replicates the hyperbolic bulb structure found in the traditional Mandelbrot set. These simulations have been carried out using the \texttt{netbrot} \cite{netbrot_2025_15573720} application. The images were rendered at a resolution of $1200 \times 1200$ with a maximum of 8192 iterations until the point is considered to be in the equi-M set. The periodicity $k$ was determined by computing the $k$ shifted difference and comparing to a tolerance of $10^{-3}$. We identified disjoint subsets of the equi-M set characterized by different post-critical combinatorics ({\it pseudo-bulbs} ${\cal B}_2^k$). Figure~\ref{higher_periods1} illustrates these low period pseudo-bulbs for two examples. One is the simple forward network in Example 1, with iterations given by $z_1(n+1)= z_1^2+c$ and $z_2(n+1) =(z_1+z_2)^2+c$. The other is the network in Example 2(b) with $w = 0.4$ . We know that ${\cal B}_2^k$, $k \geq 1$, do not form an exhaustive partition for the equi-M (since we found examples of chaotic critical dynamics in Example 2(b)).

\begin{figure}[ht!]
\centering
\fbox{\includegraphics[width=0.4\textwidth]{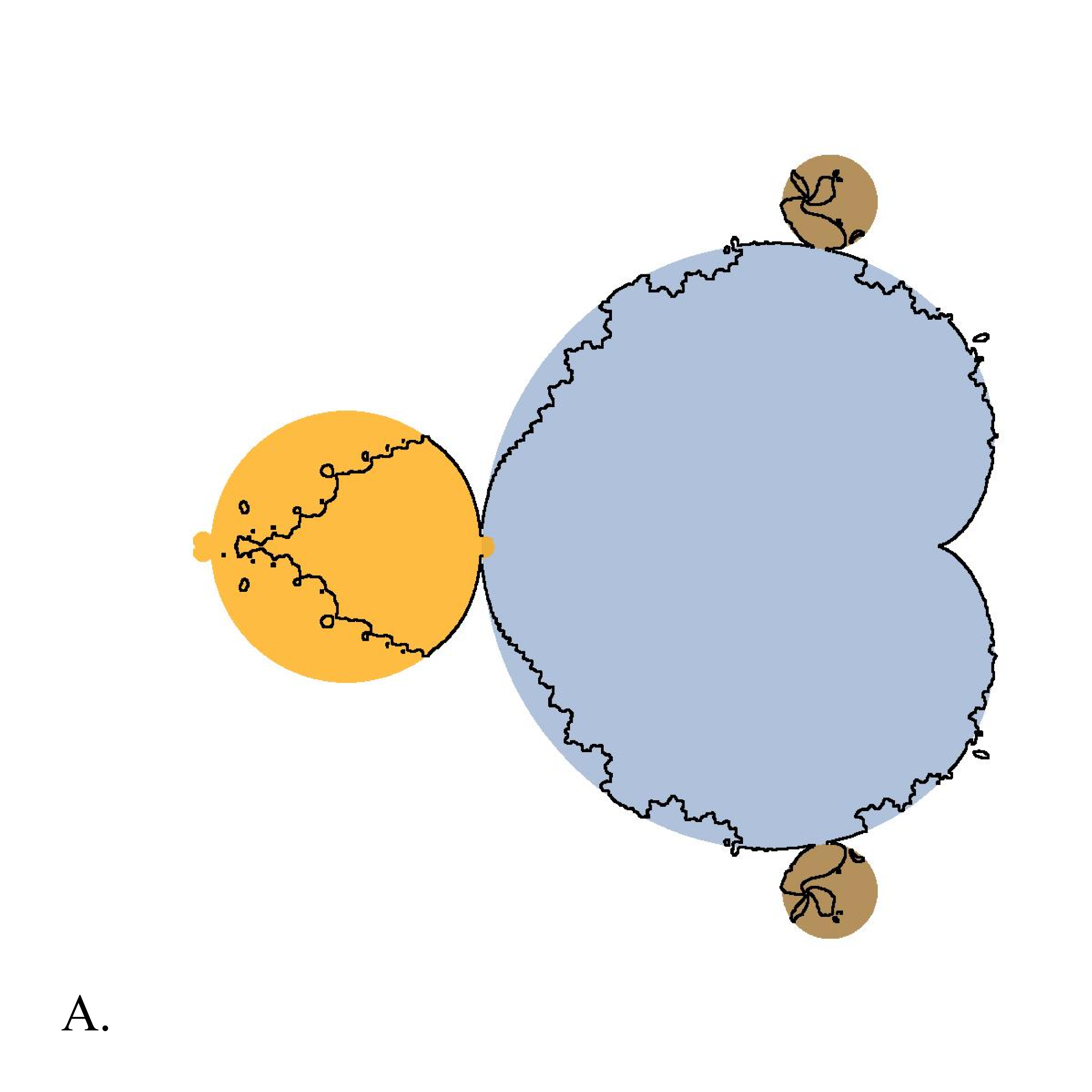}}
\quad \quad
\fbox{\includegraphics[width=0.4\textwidth]{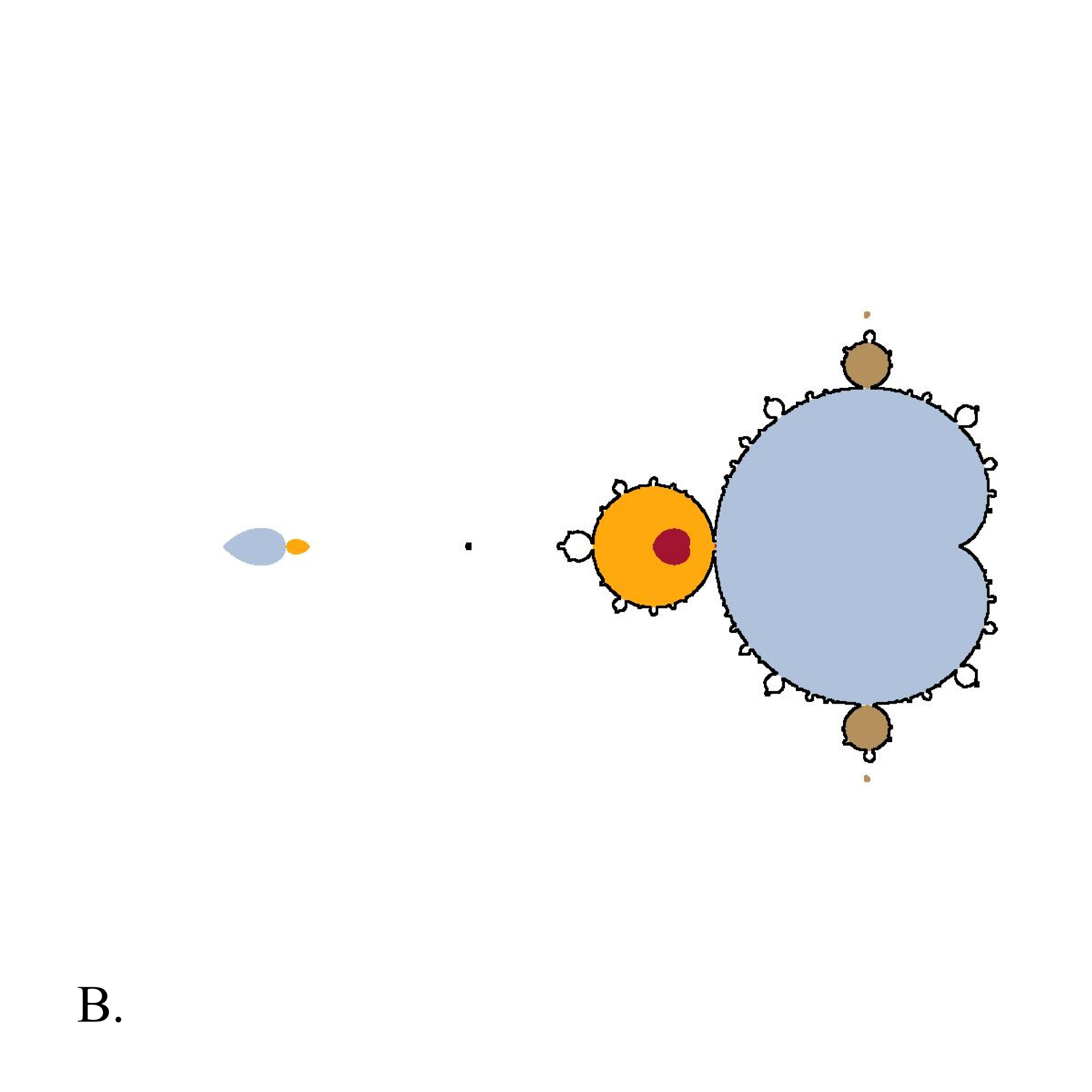}}
\caption{\small {\bf Regions with higher order periods for two example systems:} {\bf A.} The feed-forward network $z_1 = z_1^2+c$; $z_2=(z_1+z_2)^2+c$ and {\bf B.} The system in Example 2(b), for $\omega = 0.4$. In each panel, the shaded regions correspond to the system having an attracting fixed point (blue), an attracting period 2 orbit (yellow) and a period 3 orbit (green). In the right panel, the brown region represents coexistence of an attracting fixed point and an attracting period two orbit. In both cases, the boundary of the equi-M set is shown as a black contour, for comparison.}
\label{higher_periods2}

\end{figure}

For both example systems, the boundary of the main equi-cardioid ${\cal C}_2^1$ was already computed analytically and illustrated in Section \ref{cardioid}. Based on the information in Figure~\ref{higher_periods2}, we see that, unlike in the case of single quadratic map iterations, the main cardioid does not simply identify in either case with the boundary of the period $k = 1$ pseudo-bulb (purple region in the  Figure~\ref{higher_periods1} examples). For the system in Example 1, there are points where an attracting fixed point exists (i.e., $c$ is inside of the main cardioid, shown in Figure~\ref{higher_periods2}a as the blue shaded region), yet the critical point escapes (i.e., $c$ is outside of the black equi-M contour). The situation is even more interesting in Example 2, where the $c$-locus inside the main cardioid (where the system has an attracting fixed point, shaded in blue in Figure~\ref{higher_periods2}b)  is composed of three connected components: the rightmost (and largest) lies completely inside the equi-M contour; the leftmost is completely outside of the equi-M contour; the middle component is in the equi-M set, and it is a subset where an attracting fixed point and an attracting period two coexist (brown region).

This phenomenon is important, since it indicates a departure from the results in the single iterated map case. It is well-known that, for polynomial families, the critical orbits encompass global information on all other possible orbits. In particular, this is true for the complex quadratic family, with the  Mandelbrot set offering a comprehensive atlas of all combinatorics. Our two examples clarify that this is no longer the case 2D-CQNs, where one cannot expect the postcritical behavior to be accurately and completely descriptive of the whole system dynamics. More implications of this extension are contextualized in Section \ref{discussion}.

\section{Synchronization in systems of two coupled maps}
\label{synchronization}

The structure of the equi-M set of \eqref{mother_sys} is determined by the intersection of node-wise equi-M sets:

\begin{defn}
For each node $z_k$, $1 \leq k \leq n$, we define the node-wise equi-M set $\mathcal{M}(z_k)$ as the parameter locus $c \in \mathbb{C}$ for which the component corresponding to the node $z_k$ of the critical multi-orbit is bounded.
\end{defn}

In previous work on CQNs~\cite{Ariel,AC2}, we showed that nodes can have independently different behaviors, where some node clusters remain bounded, while other node clusters asymptotically escape to infinity. The number and formation of these clusters is related to network architectural structures (such as reaches or strong components), but the exact correspondence is hard to address in general networks. Two-dimensional CQNs are an excellent family of examples for anchoring basic conditions for synchronization. In this section, we build a sequence of results that track the impact of coupling parameters on the synchronization and de-synchronization of the two nodes, as well as the phase transitions between them.

\begin{lemma} \label{basic_lemma}
If the edge from $z_2$ to $z_1$ has nonzero weight, and if $\lvert z_1(n) \rvert$ is bounded as $n \to \infty$, then $\lvert z_2(n) \rvert$ is also bounded.
\end{lemma}

\begin{proof}
Assume that $z_1$ is bounded and that the coupling weight $b$ from $z_2$ to $z_1$ is nonzero. Then, we prove that $z_2$ is also bounded. Since $z_1$ is bounded, there exists an $ M > 0$, such that $|z_1(n)| \leq M$, for all $n \geq 0$. Since $|z_1(n + 1) | \geq |a z_1 + b z_2|^2 - |c|$, it follows that $|b z_2| - |a z_1| \leq \sqrt{|z_1(n + 1)| + |c|}$. Then,
\[
|b z_2 | \leq \sqrt{|z_1(n + 1)| + |c|} + |a z_1| \leq \sqrt{M + |c|} + |a| M.
\]

Then, if $b \neq 0$, we have, for all $n \geq 1$
\[
|z_2(n)| \leq K = \frac{\sqrt{M + |c|} + |a | M}{|b|}.
\]
\end{proof}

\noindent {\bf Note:} Analogously, we can show that, assuming $z_2$ is bounded and the coupling weight
$d$ from $z_1$ to $z_2$ is nonzero, $z_1$ remains bounded as well. Then we have the following:

\begin{corol}
\label{z1z2_bounded}
If the two nodes $z_1$ and $z_2$ are connected both ways by edges with nonzero weight, then the nodes are simultaneously bounded.
\end{corol}

In previous work~\cite{Ariel,AC2}, we defined ``synchronization'' of the components of the critical orbit into different clusters, based on bounded versus unbounded behavior. More precisely:

\begin{defn}
We say that the two network nodes $z_1$ and $z_2$ are M-synchronized, if their equi-M sets $\mathcal{M}(z_1)$ and $\mathcal{M}(z_2)$ are identical. We also say that the nodes belong to the same M-synchronization cluster (or simply M-cluster) of the network.
\end{defn}

In synchronization terms, Corollary~\ref{z1z2_bounded} tells us that, if $b,d \neq 0$, then the nodes are synchronized and $\mathcal{M}(z_1)=\mathcal{M}(z_2)$.

\begin{figure}[h!]
\centering
\includegraphics[width=0.55\textwidth]{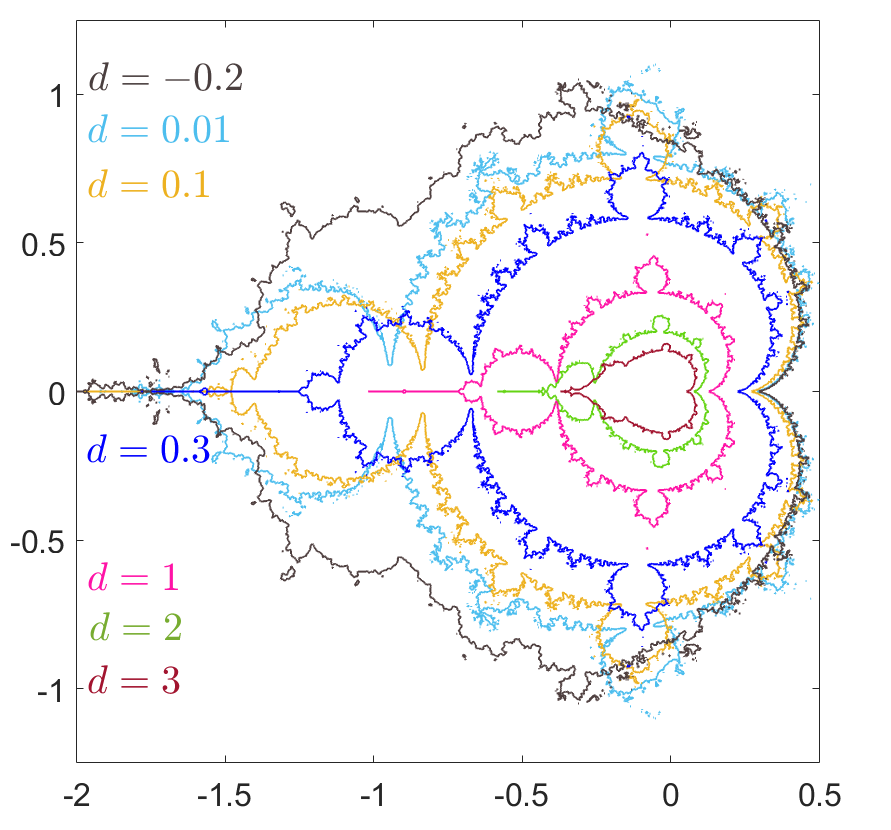}

\caption{{\small {\bf Evolution of the equi-M set as the value of $d$ increases.} The other entries of the matrix $A$ were fixed to $a=0.6$, $b=0.8$, $f=0.4$. The panel shows the equi-M contour (identical between the two nodes), using different colors for different values of $d$, as follows: $d=-0.2$ (grey); $d=0.01$ (cyan); $d=0.1$ (orange); $d=0.3$ (blue, scaling of the traditional Mandelbrot set, since $d$ is the critical values for which $\det(A)=0$); $d=1$ (pink); $d=2$ (green); $d=3$ (brown). The equi-M sets were all computed in resolution $800 \times 800$, for 50 iterations and escape radius $R=20$ (chosen large enough to apply to all cases, based on Theorem~\ref{thm.escape.radius}).}}
\label{d_all}
\end{figure}

Furthermore, if the weight matrix $A$ is singular and $a,b \neq 0$, \eqref{mother_sys} reduces to a scalar system. This can be achieved by taking $\xi = (a+b k^2)(a z_1 + b z_2$, such that $\xi(n + 1) = \xi^2+(a + b k^2) (a + b) c$, where $k = \frac{d}{a} = \frac{f}{b}$. The values of $c$ for which $z_1$ and $z_2$ are simultaneously bounded are the same as the values of $c$ for which $(a + b k^2) (a + b) c$ is in the traditional Mandelbrot set. Hence, the equi-M set of~\eqref{mother_sys} (identical between the two nodes) is a rescaling of the traditional Mandelbrot set. Figure~\ref{d_all} shows how the shape of the equi-M set of the two synchronized nodes evolves as the value of $d$ increases through the negative and positive range, while the other three entries of $A$ remain fixed. Notice that, as $d$ increases from negative values, the shape slowly approaches that of the traditional Mandelbrot set (achieved for the critical $d = 0.3$) and then slowly degrades away from this shape as $d$ continues to increase.

This sheds some light on the effects of the coupling when the nodes are interconnected ($b, d \neq 0$). The opposite case, where the nodes both act independently, is
trivially described by the following result.

\begin{lemma}
If $b = d = 0$ and $a, f \neq 0$, then the two node-wise equi-M sets are both scaled versions of the traditional Mandelbrot set.
\end{lemma}

\begin{proof}
We have $z_1(n+1) = (az_1)^2+c$, and $z_2(n+1) = (fz_2)^2+c$. Then $c \in \mathcal{M}(z_1) \iff ac \in \mathcal{M} \iff \mathcal{M}(z_1) = \frac{1}{a^2}\mathcal{M}$. Analogously, $\mathcal{M}(z_2) = \frac{1}{f^2}\mathcal{M}$.
\end{proof}

Interestingly, the two nodes in a 2D-CQN do not need to be nontrivially interconnected in order to be M-synchronized. The following lemma gives a weaker condition than Corollary~\ref{z1z2_bounded}, that ensures M-synchronization of the two nodes.

\begin{lemma} \label{weak_f}
Let $z_1(n + 1) = (a z_1)^2 + c$ (decoupled) and $z_2(n + 1)=(d z_1 + f z_2)^2 + c$ (coupled). Assume that $|f|$ is sufficiently small. Then, if the $z_1$ component of the critical orbit is bounded, the $z_2$ component is also bounded.
\end{lemma}

\begin{proof}
Since $z_1$ is bounded, there exists a large enough $M > 0$ such that $|z_1(n)| \leq M$, for all $n \geq 0$. The value of $M$ depends on $a$ and $|c|$. Assume that $|f|$ is small enough to satisfy
\begin{equation} \label{eq:f_bound}
4 M |f d| + 4 |f|^2 |c| < 1.
\end{equation}

Then, the discriminant of the quadratic function
\[
h(X) = |f|^2 X^2 + (2 M |d f| - 1) X + M^2 |d|^2 + |c|
\]
is $\Delta = 1 - 4 M |f d| - 4|f|^2 |c| > 0$. Let $X_1 < X_2$ be the two distinct roots of $h(X) = 0$. Then, the larger root $X_2$ is positive. Take $K$ such that $\max \{X_1, 0\} < K < X_2$. Clearly, we have that $h(K) < 0$. We show inductively that $K$ is an upper bound for $z_2(n)$, for all $n \geq 0$. Indeed, $z_2(0) = 0 < K$. Now suppose $|z_2| = |z_2(n)| < K$, and consider
\[
\begin{aligned}
|z_2(n+1) |
    & = |(d z_1 + f z_2)^2 + c | \leq |d z_1 + f z_2 | ^2 + |c| \\
    & \leq (|d | M + |f z_2 | )^2 + |c | \\
    & = |d|^2 M^2 + 2 |f d| M |z_2| + |f|^2 |z_2|^2 + |c| \\
    & \leq |d|^2 M^2 + 2|f d| M K + |f|^2 K^2 + |c| \\
    & = h(K) + K < K.
\end{aligned}
\]

Hence $|z_2(n+1) | < K$, and the induction is complete. In conclusion, $K$ is an upper bound for the critical component $z_2$.
\end{proof}

\begin{corol} \label{f_threshold}
If $z_1(n+1) = (az_1)^2 + c$, and $z_2(n+1)=(dz_1+fz_2)^2+c$, with $a,d \neq 0$ and $|f |$ sufficiently small, then $\mathcal{M}(z_1) = \mathcal{M}(z_2)$.
\end{corol}

\begin{proof}
Follows from Lemmas~\ref{basic_lemma} and~\ref{weak_f}.
\end{proof}

\begin{remark}
Corollary~\ref{f_threshold} states that, if node $z_2$ depends on $z_1$, but $z_1$ does not depend on $z_2$, then a sufficiently weak self-dependence of $z_2$ will still ensure M-synchronization. The threshold magnitude of self-dependence $f$ is defined by~\eqref{eq:f_bound}.
\end{remark}

\begin{figure}[h!]
\centering
\includegraphics[width=\textwidth]{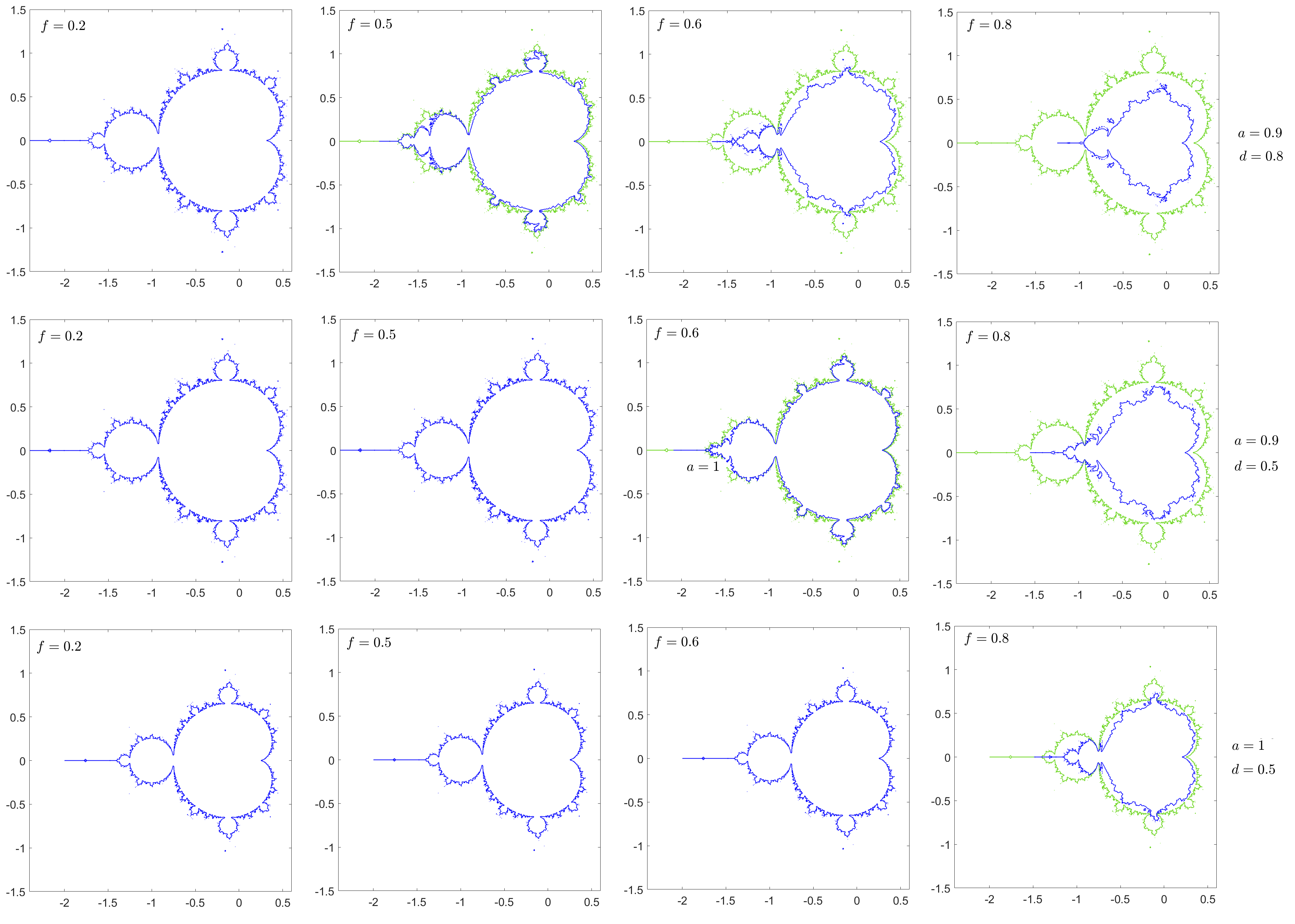}
\caption{\small {{\bf Equi-M Evolution of M synchronization in a 2D CQN when changing $f$.} Each panel shows the two contours of the node-wise equi-M sets, in green (for $z_1$) and in blue (for $z_2$), for the parameter values shown in each case. The equi-M sets were all computed in resolution $800 \times 800$, for 50 iterations and escape radius $R=20$ (chosen large enough to apply to all cases).}}
\label{evolution_f}

\end{figure}

Figure~\ref{evolution_f} shows the behavior of the equi-M contours for positive values of $f$, $b = 0$ and $a, d \neq 0$. We can see that the nodes eventually synchronize in each case for small enough values of $|f|$. In addition, as shown in Corollary~\ref{f_threshold}, the threshold depends on the values of $a$ and $d$.

\section{Discussion}
\label{discussion}

This paper continues our investigation of CQNs by returning to the foundational aspects of the framework and revisiting basic questions. This marks a departure from our prior work, which has been primarily focused on understanding the relationship between network architecture and emerging dynamic behavior (as captured by the equi-M set). Through a combination of analytical and numerical approaches, we sought to examine how different coupling architectures affect the long-term behavior of the system, and how these effects are reflected in the geometry of the equi-M set, in both low- and high-dimensional networks.

In particular, we explored the extent to which the equi-M set captures meaningful information about the global asymptotic dynamics of the system. This is easier to do in low-dimensional coupled systems, which offer higher analytical tractability. Systems with fewer nodes exhibit simpler architectures, making it possible to derive general principles that characterize emerging dynamics in CQNs, rather than the dependence of behavior on specific architectural features. This contrasts with our previous studies, where we prioritized understanding the effects of network structure, even when working with low-dimensional systems.

One central question we approached is to which extent the equi-M set of 2D CQNs replicates the hyperbolic bulb structures seen in single-variable iterated maps. To this end, we examined the parameter regions in which the critical orbit converges to an attractor of period
$k \geq 1$. To identify these periodic attractors and track the behavior of the critical orbit, we used analytical techniques for the period-1 case and numerical methods for higher periods. We then assessed whether the dynamics of the critical orbit remain representative of the full range of admissible combinatorics in the system.

Our current findings show that, unlike the simple family of one-dimensional quadratic maps, coupled systems may support coexisting local attractors of identical or differing periods. Remarkably, even in the presence of such attractors, the critical orbit can escape to infinity in $\mathbb{C}^2$. This undermines any attempt to establish a comprehensive classification of postcritical combinatorics, in any simple way that would be comparable with the partition into hyperbolic bulbs provided for the one-dimensional setting.

To further investigate this phenomenon, we analyzed more carefully the parameter regions where the critical orbit converges to an attracting fixed point -- what we called the {\it main equi-cardioid}. We conducted explicit computations across a broad range of one-parameter families, showing how specific parameter choices influence the geometry of the main equi-cardioid and shape key topological features such as connectedness and the presence of a right-hand cusp. Fast numerical computation of the parameter region ${\cal B}_2^1$, were the critical orbit converges to a fixed point, reveals that this region aligns with the boundary of the equi-cardioid ${\cal C}_2^1$ near the rightmost point on the real axis (typically corresponding to the cusp). However, ${\cal B}_2^1$ misses regions inside the equi-cardioid where attracting fixed points exist, but the critical orbit still escapes. This discrepancy persists for higher-period attractors as well. Nevertheless, the close alignment of the \sout{period-1 pseudo-bulb} the ${\cal B}_2^1$ locus and the equi-cardioid near the cusp make the boundary of the equi-M set a reasonable approximation for both these structures in this region. As one moves away from the cusp, however, this alignment breaks down: the postcritically bounded set no longer provides a faithful geometric approximation of the loci corresponding to higher-period dynamics. While this result is somewhat disappointing, it is not unexpected, given the greater complexity introduced by coupling. A more complex question to raise is whether the density of hyperbolicity hypothesis becomes easier to disprove in the CQN multi-dimensional context (starting with two coupled nodes, but also for higher-dimensional networks). While we are considering this line of questioning in our current work, it is not within the scope of this paper.

Importantly, our results support the idea that the geometry of the equi-M set near the main cusp offers a better predictor of the system’s global combinatorics than other regions of the set. This is an encouraging find, as our earlier work linked variations in network coupling to the position and orientation of the main cusp. For instance, in our analysis of large brain networks, we observed that the main cusp could occur on either side of the equi-M set~\cite{Simone}, and follow-up studies confirmed that such shifts can be observed in theoretical networks of size greater or equal to three~\cite{AC2}, suggesting that the shift is influenced by the number and placement of negative weights in the network. The current result reinforces the idea that such geometric features of the equi-M set reflect meaningful aspects of the network’s underlying architecture and its associated combinatorics.

The study of two-dimensional systems also provides new insight into the phenomenon of node synchronization, allowing us to identify explicit conditions for node-wise components of the equi-M set to be identical. This is a topic we have explored in earlier work. While synchronization is likely a key factor in the function of complex dynamical networks, it is difficult to analyze directly in large-scale systems such as brain networks. In our previous studies, we proposed general principles for how synchronization may arise, but it remained challenging to identify specific architectural patterns that promote it. The two-node case offers a tractable model in which only two dynamical outcomes are possible (the nodes are either synchronized or not), allowing us to identify key coupling parameters and analytically compute the conditions for transitions between synchronized and desynchronized behavior. These findings are valuable not only in their own right but also as a demonstration of how low-dimensional systems can yield foundational insights that are both scalable and generalizable. Our ongoing work aims to identify necessary and sufficient conditions that govern synchronization in three-node systems, where the complexity increases significantly due to added degrees of freedom and a richer combinatorial structure.

Overall, our study refines the role of the equi-M set as a novel and rigorous analytic tool for visualizing and classifying how coupling reshapes the parameter space of complex quadratic systems. Our results extend the intuition of the traditional Mandelbrot set to systems of two interacting maps. The two-dimensional case reveals, in concrete analytical form, how even minimal coupling produces new geometric and dynamical structures, offering insight into the behavior of larger networks.

\subsection*{Acknowledgements}

The project received support from an AMS-Simons PUI Faculty Grant (R\u{a}dulescu) and from a joint NSF-UEFISCDI grant: Project No. \#2408407 (R\u{a}dulescu), Project No. ROSUA-2024-0002 (Kaslik, Fikl).

\section*{Appendix A}

For the family of feed-forward two-dimensional coupled maps in Example 1, defined (for $f > 0$) by the iteration:
$z_1(n+1) = z_1^2 + c$ and $z_2(n+1) = (z_1 + f z_2)^2 + c$, we want to establish the boundary of the the region in the $c$-plane where the two-dimensional system has at least one stable fixed point $(z_1^*,z_2^*)$. This encompasses fixed point conditions:
\begin{align}
z^*_1 &= {z^*_1}^2+c
\tag{F1} \label{fixed1}\\
z^*_2 &= (z^*_1+fz^*_2)^2+c
\tag{F2} \label{fixed2}
\end{align}
and stability conditions on the magnitude of the eigenvalues:
\begin{equation}
2f |z^*_1+fz^*_2 | =1 \quad \text{and} \quad |2z^*_1 | \leq 1
\tag{S1} \label{stab1}
\end{equation}
\begin{equation}
|2z^*_1 | = 1 \quad \text{and} \quad 2f|z^*_1+fz^*_2 | \leq 1
\tag{S2} \label{stab2}
\end{equation}

Here too, we give the parametric description of the boundary of ${\cal C}_2^1$ in polar form. For simplicity, for the rest of this computation we drop the $(*)$ and refer to the fixed point as $(z_1,z_2)$, since there is no danger of confusion. If we eliminate $c$ from the fixed point equations, we get that $z_2-z_1 = 2fz_1z_2+f^2z_2^2$. Hence, one can express

\begin{eqnarray*}
& & z_1
    = \frac{z_2 (1 - f^2 z_2)}{1 + 2 f z_2}, \\
& & z_1 + f z_2
    = \frac{z_2 (1 - f^2 z_2)}{1 + 2 f z_2} + f z_2
    = \frac{z_2 (1 + f + f^2 z_2)}{1 + 2 f z_2}, \\
& & c = z_1 - z_1^2
    = \frac{z_2 (1 - f^2 z_2)}{1 + 2 f z_2}
    - \frac{z_2^2 (1 - f^2 z_2^2)^2}{(1 + 2 f z_2)^2}
    = \frac{z_2 (1 - f^2 z_2)[1 + (2 f - 1) z_2 + f^2 z_2^2]}{(1 + 2 f z_2)^2}.
\end{eqnarray*}

The first condition in~\eqref{stab1}  can then be rewritten as:
\[
2 f |z_2| |1 + f + f^2 z_2| = |1 + 2 f z_2|.
\]

If one considers $z_2$ in its polar form $z_2 = \rho e^{i\theta}$, the condition becomes:
\[
\begin{aligned}
&
    2 f \rho |1 + f + f^2 \rho e^{i \theta}| = |1 + 2 f \rho e^{i\theta}| \\
\iff\,\, &
    4 f^2 \rho^2 [(1 + f)^2 + 2 f^2 (1 + f) \rho \cos \theta + f^4 \rho^2]
    = 1 + 4 f \rho \cos \theta + 4 f^2 \rho^2.
\end{aligned}
\]

Solving for $\theta$, we obtain
\begin{equation} \label{S1a}
\cos(\theta) =
    \frac{1 - 8 f^3 \rho^2 - 4 f^4 \rho^2 - 4 f^6 \rho^4}{8 f^4 \rho^3 (1 + f) - 4 f \rho}
     \tag{S1a}
\end{equation}

In turn, the second condition in~\eqref{stab1} becomes
\[
\begin{aligned}
&
    2 \rho |1 - f^2 \rho e^{i\theta} |
    \leq |1 + 2 f \rho e^{i\theta}| \\
\iff\,\, &
    4 \rho^2 (1 - 2 f^2 \rho \cos \theta + f^4 \rho^2)
    \leq 1 + 4 f \rho \cos \theta + 4 f^2 \rho^2.
\end{aligned}
\]

If one calls
\begin{equation}
h(\rho) = \frac{4 f^4 \rho^4+4\rho^2(1-f^2)-1}{4f\rho(2f\rho^2+1)},
\end{equation}
then condition in polar form becomes:
\begin{equation}
\cos \theta \geq h(\rho). \tag{S1b} \label{S1b}
\end{equation}

To interpret conditions~\eqref{S1a} and~\eqref{S1b} together, we distinguish the following three cases:

\begin{description}
\item[Case I:] $h(\rho) > 1$, so that there is no angle $\theta$ that satisfies~\eqref{S1b}. The condition on $h$ is equivalent to
\[
    4 \rho^2 (f^2 \rho - 1)^2 > (2 f \rho + 1)^2.
\]

\begin{itemize}
    \item If $f^2 \rho > 1$, then this is further equivalent to asking that $2 f^2 \rho^2 - 2 \rho (f + 1) - 1 > 0$. Since the two quadratic roots
\[
K^{\pm}(f) = \frac{1 + f \pm \sqrt{(f + 1)^2+2f^2}}{2f^2}
\]
are such that $K^-(f) < 0 < f^{-2} < K^+(f)$ for all $f > 0$, we get that this scenario occurs when $\rho > K^+(f)$.

    \item If $f^2 \rho < 1$, then the condition becomes $f^2\rho + 2\rho(f-1)+1<0$. This has no solutions for $\rho$ when $f>\sqrt{2}-1$. When $f<\sqrt{2}-1$, the quadratic function has two roots
\[
M^{\pm}(f) = \frac{1-f \pm \sqrt{(1-f)^2-2f^2}}{2f^2}.
\]

Since $f < 1$, the roots are such that $0 < M^-(f) < M^+(f) < f^{-2}$, hence this scenario occurs only for $f < \sqrt{2}-1$, when $M^-(f) < \rho < M^+(f)$.
\end{itemize}

Notice that the two intervals for $\rho$ (where there is no angle $\theta$ that satisfies~\eqref{S1b}) are disjoint, since $M^+(f)<\frac{1}{f^2}< K^+(f)$.

\item[Case II:] $h(\rho) < -1$, hence~\eqref{S1b} generates no additional conditions on $\theta$. The condition is equivalent to
\[
4\rho^2(f^2\rho +1)^2 < (2f\rho-1)^2.
\]

\begin{itemize}
    \item If $2 f \rho > 1$, then this is further equivalent to $2f^2\rho^2+2(1-f)\rho+1<0$. When $f>\sqrt{2}-1$, the quadratic roots are complex, and there are no solutions for $\rho$. When $f<\sqrt{2}-1$, both roots
\[
    R^{\pm}(f) = \frac{f - 1 \pm \sqrt{(1 - f)^2 - 2 f^2}}{2 f^2} < 0,
\]
hence there is no solution here, either.

    \item If $2 f \rho < 1$, then the condition becomes $2f^2\rho^2+2(1+f)\rho-1<0$, so that there are always two real roots
\[
S^{\pm}(f) = \frac{-(f+1)\pm \sqrt{(f+1)^2+2f^2}}{2f^2}.
\]

In addition, we notice that $S^-<0<S^+<\frac{1}{2f}$.
\end{itemize}

We continue investigating the $2 f \rho < 1$ case that allows valid solutions.
If $0<\rho<S^+$, the additional restriction coming from~\eqref{S1b} is that $-1 \leq \cos \theta \leq 1$, hence:
\[
-1 \leq 1-8f^3 \leq 1
\]
and
\[
\ds \rho < \frac{-2+\sqrt{6}}{2}
    \iff 2\rho^2+4\rho-1<0
    \iff 2\rho^2-1<-4\rho
    \iff h(\rho)<-1.
\]

Hence in this case condition~\eqref{S1b} is satisfied by all values of $\theta$.

Since $\ds \rho < \frac{-2+\sqrt{6}}{2} <\frac{1}{2}$, then $4\rho^2-1<0$. Using the expression in~\eqref{S1a} for $\cos \theta$ we get that the only condition on the angle is that $-1 \leq \cos \theta \leq 1$, which can be rewritten as:
\begin{equation}
-16\rho^4+4\rho \geq 1-12\rho^2-4\rho^4.
\tag{IIa} \label{IIa}
\end{equation}
and
\begin{equation}
1-12\rho^2-4\rho^4 \geq 16\rho^3-4\rho.
\tag{IIb} \label{IIb}
\end{equation}

One can calculate that~\eqref{IIa} is equivalent to:
\begin{equation*}
4\rho^4-16\rho^3+12\rho^2+4\rho-1 \geq 0 \iff 4\rho^2(\rho-2)^2 \geq (2\rho-1)^2.
\end{equation*}

Since in this case $\rho<1/2<2$, it follows that~\eqref{IIa} is equivalent with the quadratic inequality $2\rho(2-\rho) \geq 1-2\rho$, which is satisfied when
\begin{equation}
\frac{3-\sqrt{7}}{2} < \rho <\frac{3+\sqrt{7}}{2}
\tag{IIa$'$} \label{IIa'}
\end{equation}

The second inequality~\eqref{IIb} is equivalent to:
\begin{equation*}
4\rho^4+16\rho^3+12\rho^2-4\rho-1 \leq 0 \iff 4\rho^2(\rho+2)^2 \geq (2\rho+1)^2
\end{equation*}

This is equivalent with the quadratic inequality $2\rho(2+\rho) \geq 1+2\rho$, which is satisfied when
\begin{equation}
0 < \rho <\frac{-1+\sqrt{3}}{2}
\tag{IIb$'$} \label{IIb'}
\end{equation}

Since $\ds \frac{-1+\sqrt{3}}{2}>\frac{-2+\sqrt{6}}{2}$ and $\ds \frac{3+\sqrt{7}}{2}>\frac{-\sqrt{6}}{2}$, the two conditions~\eqref{IIa'} and~\eqref{IIb'} combined lead to the interval:
 \begin{equation}
\frac{3-\sqrt{7}}{2} < \rho <\frac{-2+\sqrt{6}}{2}
\tag{II$'$} \label{II$'$}
\end{equation}

\item[Case III:] $\ds \frac{-2+\sqrt{6}}{2} \leq \rho \leq \frac{2+\sqrt{6}}{2} \iff -1 \leq h(\rho) \leq 1$. In this case we have nontrivial solutions for $\theta$, so we need to consider both conditions~\eqref{IIa} and~\eqref{IIb} when computing the domain for $\rho$, as follows:
\begin{equation*}
h(\rho) = \frac{2\rho^2-1}{4\rho} \leq \frac{1-12\rho^2-4\rho^4}{4\rho(4\rho^2-1)} \leq 1
\end{equation*}

We distinguish two subcases:

\begin{itemize}
\item If $\ds \rho < \frac{1}{2}$. Then we have~\eqref{IIb'} and also (since $4\rho^2-1<0$)
\[
(2\rho^2-1)(4\rho^2-1) \geq 1-12\rho^2-4\rho^4 \iff 12\rho^4+6\rho^2 \geq 0
\]
which is satisfied by all $\rho$. Since $\ds \frac{-1+\sqrt{3}}{2}<\frac{1}{2}$, the conditions for $\rho$ can be written as
\[
\frac{-2+\sqrt{6}}{2} \leq \rho \leq \frac{-1+\sqrt{3}}{2}.
\]

\item If $\ds \rho > \frac{1}{2}$. Then $4\rho^2-1>0$ and we have
\begin{equation*}
(2\rho^2-1)(4\rho^2-1) < 1-12\rho^2-4\rho^4 < 16\rho^3-4\rho \iff 12\rho^2 + 6\rho^2  <0
\end{equation*}
which has no solution.
\end{itemize}

In conclusion, Case III allows for
\[
\frac{-2+\sqrt{6}}{2} \leq \rho \leq \frac{-1+\sqrt{3}}{2}.
\]

\end{description}

This closes the discussion of the joint conditions~\eqref{S1a} and~\eqref{S1b}, so that~\eqref{stab1} can be altogether translated into the following polar curve, defined parametrically as:
\begin{equation}
\cos \theta = \frac{1-12\rho^2-4\rho^4}{4\rho(4\rho^2-1)}, \text{ for } \frac{3-\sqrt{7}}{2} \leq \rho \leq \frac{-1+\sqrt{3}}{2}
\tag{S1$'$} \label{S1'}
\end{equation}

No additional boundary points result from conditions~\eqref{stab2}. Indeed, considering again the polar form for $z_2 = \rho e^{i\theta}$ we can rewrite the first part of~\eqref{stab2} as:
\begin{equation*}
|z_1 | = \left |\frac{z_2(1-z_2)}{1+2z_2}  \right | = \frac{1}{2}\\ \\ \iff 2\rho \sqrt{1-2\rho \cos \theta + \rho^2}  = \sqrt{1+4\rho \cos \theta + 4\rho^2}
\end{equation*}
leading to
\begin{equation}
\cos \theta = \frac{4\rho^4-1}{4\rho (2\rho^2+1)} = \frac{2\rho^2-1}{4\rho}
\tag{S2a} \label{S2a}
\end{equation}

The second part of~\eqref{stab2} becomes:
\begin{equation*}
2|z_1+z_2 | = \frac{2|z_2 | |2+z_2 |}{\vert 1+2z_2 |} =1 \iff 2\rho \sqrt{4+4 \rho \cos\theta+\rho^2} \leq \sqrt{1+4\rho \cos\theta+4\rho^2}
\end{equation*}
from which it follows that
\begin{equation*}
1-12\rho^2-4\rho^4 \geq 4\rho (4\rho^2-1) \cos \theta = 4\rho (4\rho^2-1) \frac{2\rho^2-1}{4\rho} = 8\rho^4-6\rho^2+1
\end{equation*}

Hence $12\rho^4+6\rho^2 \leq 0$, which produces no valid solutions for $\rho$. This concludes the proof that the conditions in~\eqref{stab2} are empty, and the bounding curve of the region with one locally attracting fixed point is indeed given by the expression in~\eqref{S1'}, as represented in Figure~\ref{fig:cardio.feedforward.1}.

\section*{Appendix B}

Consider the family of coupled maps subject to the condition $a + b = d + f = \omega$. Then equation \eqref{eq.u1.u2} provides:
\[
(u_1-u_2)\left[(a-d)(u_1+u_2)-1\right]=0,
\]
where, as before, $u_1=az_1+bz_2$ and $u_2=dz_1+fz_2$. We have two cases:\\

\begin{description}
\item[Case I:] $u_1=u_2$. In this case, system \eqref{eq.fixed.point} reduces to:

\begin{equation} \label{eq.fixed.point.w.case1}
c\omega +\omega u_1^2=u_1
\end{equation}
and the characteristic equation becomes:
\[
\lambda^2 - 2 \tau u_1 \lambda + 4 \delta u_1^2 = 0,
\quad \text{where } \tau = \tr(A) = a + f ~\text{and}~\delta=\det(A)=af-bd,
\]
or, equivalently,
\[
\left(\frac{\lambda}{2 u_1}\right)^2
- \tau \left(\frac{\lambda}{2 u_1}\right)
+ \delta = 0,
\]
which means that $\lambda / (2 u_1) \in \sigma(A) = \{\omega, \tau - \omega\}$. Therefore, eliminating $u_1$ from \eqref{eq.fixed.point.w.case1}, it follows that:
\[
\lambda^2 - 2 \frac{\mu}{\omega} \lambda + 4 c \mu^2 = 0,
\quad \text{where } \mu \in \{\omega,\tau-\omega\}.
\]

If $\mu=\omega$, from Vieta's formulas we deduce that $\lambda_1+\lambda_2=2$. Therefore, at least one eigenvalue has absolute value larger than one, and the corresponding fixed point is unstable.

On the other hand, if $\mu = \tau - \omega$, then $\alpha = \frac{\tau}{\omega} - 1$ and $\beta = 4 c (\tau - \omega)^2$, and the inequalities from Lemma \ref{lem.quadratic.poly} become:
\begin{equation}\label{ineq.cardio.1}
|c| < \frac{1}{4 (\tau - \omega)^2}
\quad \text{and} \quad
2 \left|\frac{\tau}{\omega} - 1\right|
< \frac{1 - 16 (\tau - \omega)^4 |c|^2}
       {\sqrt{1 + 16 (\tau - \omega)^4 |c|^2 - 8 (\tau - \omega)^2 \Re(c)}}.
\end{equation}

These inequalities define a part of the cardioid only if $\left|\frac{\tau}{\omega} - 1\right| < 1$. Otherwise, the second inequality from \eqref{ineq.cardio.1} cannot hold, and consequently, both fixed points for which $u_1=u_2$ are unstable.

\item[Case II:] $u_1+u_2=(a-d)^{-1}=(\tau-\omega)^{-1}$, $\tau\neq\omega$.

Let us write
\[
q : = u_1 - u_2,
\qquad
u_1 = \frac{1}{2} \left(\frac{1}{\tau - \omega} + q\right),
\qquad
u_2 = \frac{1}{2} \left(\frac{1}{\tau - \omega} - q\right),
\qquad q \in \mathbb{C}.
\]

Substituting the above expressions into the first equation of~\eqref{eq.fixed.point} we obtain
\begin{equation}\label{eq.c.case2}
c(q) = \frac{-a (q + 1)^{2}-b (q - 1)^2 + 2 (\tau - \omega) (q + 1)}
            {4 \omega (\tau - \omega)} .
\end{equation}

With the same parametrization we have
\[
\alpha(q) = a u_1 + f u_2
          = \frac{(\tau-\omega) q + d - \omega}{2 (\tau - \omega)},
\quad \text{and} \quad
\beta(q) = 4\delta u_1 u_2
         = \frac{\delta}{(\tau - \omega)^2} (1 - q^2).
\]

Based on Lemma~\ref{lem.quadratic.poly}, we define
\[
\mathcal{Q} = \biggl\{
    q \in \mathbb{C}\,:\,
    |\beta(q)| < 1 \text{ and }
    2 |\alpha(q) - \overline{\alpha(q)} \beta(q)| < 1 - |\beta(q)|^{2}
\biggr\}.
\]

The set of complex parameters $c$ for which the coupled quadratic map has a
stable fixed point satisfying $u_{1} + u_{2}=(\tau-\omega)^{-1}$ is
\[
c(\mathcal{Q}) = \bigl\{c(q)\,:\, q \in \mathcal{Q}\bigr\},
\]
where $c(q)$ is given by~\eqref{eq.c.case2}.
\end{description}

\bibliographystyle{plain}
\bibliography{references}

@article{AC2,
  title         = {Synchronization and Clustering in Complex Quadratic Networks},
  author        = {
    Anca R\u{a}dulescu and Danae Evans and Amani-Dasia Augustin and Anthony
    Cooper and Johan Nakuci and Sarah Muldoon
  },
  year          = 2023,
  journal       = {Neural Computation},
  publisher     = {{MIT} Press},
  volume        = 36,
  pages         = {75--106},
  doi           = {10.1162/neco_a_01624},
  issue         = 1,
}

@article{Ariel,
  title         = {Real and Complex Behavior for Networks of Coupled Logistic Maps},
  author        = {Anca R\u{a}dulescu and Ariel Pignatelli},
  year          = 2017,
  journal       = {Nonlinear Dynamics},
  publisher     = {Springer Science and Business Media {LLC}},
  volume        = 87,
  pages         = {1295--1313},
  doi           = {10.1007/s11071-016-3115-4},
  issue         = 2,
}

@book{brucks2004topics,
  title         = {Topics From One-Dimensional Dynamics},
  author        = {Karen M. Brucks and Henk Bruin},
  year          = 2004,
  publisher     = {Cambridge University Press},
  series        = {London Mathematical Society Student Texts},
  number        = 64,
  isbn          = 9780521838962,
}

@book{carleson1996complex,
  title         = {Complex Dynamics},
  author        = {Lennart Carleson and Theodore W. Gamelin},
  year          = 1996,
  publisher     = {Springer Science \& Business Media},
  isbn          = 9780387979427,
}

@book{douady1984etude,
  title         = {\'{E}tude dynamique des polyn\^{o}mes complexes {I} \& {II}},
  author        = {A. Douady and John H. Hubbard},
  year          = 1984,
  publisher     = {Universit\'{e} de Paris-Sud, D\'{e}p. de Math\'{e}matique},
  series        = {Publications math{\'e}matiques d'Orsay},
}

@book{douady1984exploring,
  title         = {Exploring the {Mandelbrot} set},
  author        = {Adrien Douady and John H. Hubbard},
  year          = 1984,
  publisher     = {Universit\'{e} de Paris-Sud, D\'{e}p. de Math\'{e}matique},
  series        = {Publications math{\'e}matiques d'Orsay},
}

@article{douady1985dynamics,
  title         = {On the Dynamics of Polynomial-Like Mappings},
  author        = {Adrien Douady and John H. Hubbard},
  year          = 1985,
  journal       = {Annales scientifiques de l'\'{E}cole Normale Sup\'{e}rieure},
  publisher     = {Societ\'{e} math\'{e}matique de France},
  volume        = 18,
  pages         = {287--343},
  doi           = {10.24033/asens.1491},
  issue         = 2,
}

@article{DTI,
    author = { Anca Rădulescu and Eva Kaslik and Alexandru Fikl and Johan Nakuci and Sarah Muldoon and Michael Anderson },
    doi = { 10.1063/5.0283805 },
    journal = { Chaos: An Interdisciplinary Journal of Nonlinear Science },
    number = { 9 },
    publisher = { AIP Publishing },
    title = { Fractal Geometry Predicts Dynamic Differences in Structural and Functional Connectomes },
    volume = { 35 },
    year = { 2025 }
}

@article{faskowitz2022edges,
  title         = {
    Edges in Brain Networks: {C}ontributions to Models of Structure and
    Function
  },
  author        = {Joshua Faskowitz and Richard F. Betzel and Olaf Sporns},
  year          = 2022,
  journal       = {Network Neuroscience},
  publisher     = {{MIT} Press},
  volume        = 6,
  number        = 1,
  pages         = {1--28},
  doi           = {10.1162/netn_a_00204},
}

@article{gender,
  title         = {Computing Brain Networks With Complex Dynamics},
  author        = {Anca R\u{a}dulescu and Johan Nakuci and Simone Evans and Sarah Muldoon},
  year          = 2023,
  journal       = {Neural Computing and Applications},
  publisher     = {Springer Science and Business Media {LLC}},
  volume        = 35,
  pages         = {21115--21127},
  doi           = {10.1007/s00521-023-08903-4},
  issue         = 28,
}

@inproceedings{henrici1974computational,
  title         = {Computational Complex Analysis},
  author        = {Peter Henrici},
  year          = 1974,
  journal       = {Proceedings of Symposia in Applied Mathematics},
  booktitle     = {The Influence of Computing on Mathematical Research and Education},
  publisher     = {American Mathematical Society},
  pages         = {79--86},
  doi           = {10.1090/psapm/020/0349957},
}

@book{hubbard1992local,
  title         = {
    Local connectivity of {Julia} sets and bifurcation loci: {T}hree theorems
    of {J.-C. Yoccoz}
  },
  author        = {John H. Hubbard},
  year          = 1992,
  publisher     = {Institut des Hautes \'{E}tudes Scientifiques},
  series        = {Topological Methods in Modern Mathematics},
}

@article{iniguez2020bridging,
  title         = {Bridging the Gap Between Graphs and Networks},
  author        = {Gerardo I{\~n}iguez and Federico Battiston and M\'{a}rton Karsai},
  year          = 2020,
  journal       = {Communications Physics},
  publisher     = {Springer Science and Business Media {LLC}},
  volume        = 3,
  doi           = {10.1038/s42005-020-0359-6},
  issue         = 1,
}

@book{ivancevic2007high,
  title         = {
    High-Dimensional Chaotic and Attractor Systems: {A} Comprehensive
    Introduction
  },
  author        = {Vladimir G. Ivancevic and Tijana T. Ivancevic},
  year          = 2007,
  publisher     = {Springer},
  series        = {Intelligent Systems, Control and Automation: Science and Engineering},
  volume        = 32,
  isbn          = 9781402054556,
}

@phdthesis{jung2002homeomorphisms,
  title         = {Homeomorphisms on Edges of the {Mandelbrot} Set},
  author        = {Wolf Jung},
  year          = 2002,
  school        = {Bibliothek der RWTH Aachen},
}

@book{mcmullen1994complex,
  title         = {Complex Dynamics and Renormalization},
  author        = {Curtis T. McMullen},
  year          = {1994},
  publisher     = {Princeton University Press},
  number        = 135,
  isbn          = {9781400882557},
}

@book{milnor2006dynamics,
  title         = {Dynamics in One Complex Variable},
  author        = {John W. Milnor},
  year          = 2006,
  publisher     = {Princeton University Press},
  isbn          = 9780691124872,
  edition       = 3,
}

@software{netbrot_2025_15573720,
  title         = {alexfikl/netbrot: v2025.6.0},
  author        = {Alexandru Fikl},
  year          = 2025,
  month         = jun,
  publisher     = {Zenodo},
  doi           = {10.5281/zenodo.15573720},
  url           = {https://doi.org/10.5281/zenodo.15573720},
  version       = {v2025.6.0},
}

@article{robinson2022physics,
  title         = {Physics Guided Neural Networks for Modelling of Non-Linear Dynamics},
  author        = {Haakon Robinson and Suraj Pawar and Adil Rasheed and Omer San},
  year          = 2022,
  journal       = {Neural Networks},
  publisher     = {Elsevier {BV}},
  volume        = 154,
  pages         = {333--345},
  doi           = {10.1016/j.neunet.2022.07.023},
}

@article{ma2012biological,
  title={Biological network analysis: insights into structure and functions},
  author={Ma, Xiaoke and Gao, Lin},
  journal={Briefings in functional genomics},
  volume={11},
  number={6},
  pages={434--442},
  year={2012},
  publisher={Oxford University Press}
}

@article{kahn2025network,
  title={Network structure influences the strength of learned neural representations},
  author={Kahn, Ari E and Szymula, Karol and Loman, Sophie and Haggerty, Edda B and Nyema, Nathaniel and Aguirre, Geoffrey K and Bassett, Dani S},
  journal={Nature Communications},
  volume={16},
  number={1},
  pages={994},
  year={2025},
  publisher={Nature Publishing Group UK London}
}

@article{luppi2024contributions,
  title={Contributions of network structure, chemoarchitecture and diagnostic categories to transitions between cognitive topographies},
  author={Luppi, Andrea I and Singleton, S Parker and Hansen, Justine Y and Jamison, Keith W and Bzdok, Danilo and Kuceyeski, Amy and Betzel, Richard F and Misic, Bratislav},
  journal={Nature Biomedical Engineering},
  volume={8},
  number={9},
  pages={1142--1161},
  year={2024},
  publisher={Nature Publishing Group UK London}
}

@article{whitney2004influence,
  title={The influence of architecture in engineering systems},
  author={Whitney, Daniel and Crawley, Edward and de Weck, Olivier and Eppinger, Steven and Magee, Christopher and Moses, Joel and Seering, Warren and Schindall, Joel and Wallace, David},
  journal={Engineering Systems Monograph, MIT Engineering Systems Division, March},
  year={2004}
}

@article{seabrook2021evaluating,
  title         = {Evaluating Structural Edge Importance in Temporal Networks},
  author        = {Isobel E. Seabrook and Paolo Barucca and Fabio Caccioli},
  year          = 2021,
  journal       = {EPJ Data Science},
  publisher     = {Springer Science and Business Media {LLC}},
  volume        = 10,
  doi           = {10.1140/epjds/s13688-021-00279-6},
  issue         = 1,
}

@article{shishikura1998hausdorff,
  title         = {
    The Hausdorff Dimension of the Boundary of the Mandelbrot Set and Julia
    Sets
  },
  author        = {Mitsuhiro Shishikura},
  year          = 1998,
  journal       = {The Annals of Mathematics},
  publisher     = {Jstor},
  volume        = 147,
  pages         = {225--225},
  doi           = {10.2307/121009},
  issue         = 2,
}

@article{Simone,
  title         = {Asymptotic Sets in Networks of Coupled Quadratic Nodes},
  author        = {Anca R\u{a}dulescu and Simone Evans},
  year          = 2019,
  journal       = {Journal of Complex Networks},
  publisher     = {{Oxford} University Press ({OUP})},
  volume        = 7,
  pages         = {315--345},
  doi           = {10.1093/comnet/cny021},
  issue         = 3,
}

\end{document}